\documentclass[journal]{IEEEtran}
 \usepackage{cite}

\ifCLASSINFOpdf
   \usepackage[pdftex]{graphicx}
\else
\fi
 \usepackage{amsmath, amsfonts, amssymb, amsthm}

\usepackage{flushend}

\usepackage{hyperref}
\usepackage[utf8]{inputenc}
\usepackage{textcomp}
\usepackage{newunicodechar}
\newunicodechar{−}{-}
\usepackage{pifont}

\usepackage[linesnumbered,ruled,vlined]{algorithm2e} 

\begin{document}
%
\title{Quantum Inspired Vehicular Network Optimization for Intelligent Decision Making in Smart Cities}
%
%
%

\author{Kamran~Ahmad~Awan,~\IEEEmembership{Senior Member,~IEEE}, 
      Sonia~Khan,~\IEEEmembership{Member,~IEEE}, Eman Abdullah Aldakheel, Saif Al-Kuwari,~\IEEEmembership{Senior Member,~IEEE}, and~Ahmed~Farouk,~\IEEEmembership{Senior Member,~IEEE}

\thanks{Kamran~Ahmad~Awan is with the Faculty of Information Technology \& Numerical Sciences, Department of Information Technology, The University of Haripur, Haripur 22620, Khyber Pakhtunkhwa, Pakistan (e-mail: kamranawan.2955@gmail.com, kamranahmad@ieee.org, )}

\thanks{Sonia~Khan is with the Department of Youth Affairs, Youth Office Haripur, Haripur 22620, Government of Khyber Pakhtunkhwa, Pakistan e-mail: (soniakhan1.2981@gmail.com)}
\thanks{Department of Computer Sciences, College of Computer and Information Sciences, Princess Nourah Bint
 Abdulrahman University, Riyadh, Saudi Arabia. e-mail: (Eaaldakheel@pnu.edu.sa)}

\thanks{Saif Al-Kuwari is with the Qatar Center for Quantum Computing, College of Science and Engineering, Hamad Bin Khalifa University, Doha, Qatar. e-mail: (smalkuwari@hbku.edu.qa).}

\thanks{A.~Farouk is with  the Qatar Center for Quantum Computing, College of Science and Engineering, Hamad Bin Khalifa University, Doha, Qatar and the Department of Computer Science, Faculty of Computers and Artificial Intelligence, Hurghada University, Hurghada, Egypt. e-mail:(ahmed.farouk@sci.svu.edu.eg).}

\thanks{Corresponding Author: Kamran~Ahmad~Awan (kamranahmad@ieee.org)}}

\maketitle

\begin{abstract}
Connected and automated vehicles require city-scale coordination under strict latency and reliability targets, yet many approaches optimize communications and mobility separately, degrading under outages and compute contention. We present \textbf{QIVNOM}, a quantum-inspired framework that jointly optimizes V2V/V2I communication and urban traffic control on classical edge--cloud hardware (no quantum processor required). Candidate routing--signal plans are encoded as probabilistic superpositions and are updated via sphere-projected gradients with annealed sampling to minimize a regularized cost. An entanglement-style regularizer couples networking and mobility decisions, while Tchebycheff multi-objective scalarization with feasibility projection enforces latency and reliability constraints. Robustness is enhanced through chance constraints and Lyapunov-drift control. Plans are assigned to fog nodes via entropic optimal transport, and vehicle-level CVaR micro-policies align local safety with global guidance. In METR-LA--calibrated SUMO--OMNeT++/Veins simulations over a $5\times 5$~km urban map with IEEE 802.11p and 5G NR sidelink, QIVNOM reduces the mean end-to-end latency to \textbf{57.3~ms} (\(\approx 20\%\) below the best baseline). During incidents, latency is \textbf{62~ms vs 79~ms} (\(-21.5\%\)); with RSU outages, \textbf{67~ms vs 86~ms} (\(-22.1\%\)). Packet delivery averages \textbf{96.7\%} (\(+2.3\) pts), reliability \textbf{96.7\%} overall (RSU-outage \textbf{96.8\% vs 94.1\%}), and corridor-closure travel metrics improve (ATT \textbf{12.8~min}/\textbf{33\%} vs \textbf{14.5~min}/\textbf{37\%}), demonstrating consistent gains under stress. These results position QIVNOM as a practical building block for smart-city intelligent transportation systems and connected consumer electronics, enabling safer, faster, and more reliable mobility on commodity edge--cloud infrastructure.
\end{abstract}


\begin{IEEEkeywords}
Connected Vehicles, Smart Cities, Quantum-Inspired Algorithms, V2X Communication, Fog Computing, Intelligent Transportation Systems
\end{IEEEkeywords}

%
\IEEEpeerreviewmaketitle

\section{Introduction}

\subsection{An Overview}  
\IEEEPARstart{T}{he} development of urban transportation increasingly depends on connected vehicle infrastructures that form the foundation of intelligent transportation systems (ITS) \cite{11016124}. Vehicle-to-Vehicle (V2V) and Vehicle-to-Infrastructure (V2I) communications—collectively referred to as V2X—enable dynamic vehicular networks to exchange time-sensitive data between mobile nodes and roadside units \cite{10950443}. These networks require high reliability and low latency to support dense traffic conditions and continuous mobility, particularly in urban environments \cite{10948423}. However, conventional centralized architectures face limitations in meeting these demands \cite{10915641}. Vehicular ad hoc networks (VANETs), in particular, are prone to frequent topology changes and require adaptive and fault-tolerant protocols to ensure consistent data delivery under varying network conditions \cite{10908641}.  

Beyond communication, the functionality of V2X systems is heavily dependent on the computational infrastructure. Edge and fog computing frameworks are crucial to enable low-latency processing, as they facilitate localized decision-making and reduce reliance on remote cloud servers \cite{11006883}. The deployment of fog nodes near data sources enables timely analysis, which is crucial for safety-critical applications \cite{10852542}. However, integrating distributed computing resources into a cohesive and responsive architecture presents several challenges \cite{10402120}. The scale and complexity of urban mobility systems introduce a high-dimensional optimization problem involving interrelated components such as routing, signal control, spectrum management, and computation offloading \cite{10833692}. Traditional heuristic approaches and many machine learning models often fall short in handling this complexity, particularly when real-time performance and system-wide coordination are required. Although some adaptive methods show improved local performance, their scalability and consistency in large-scale deployments remain limited \cite{10930632}.

\subsection{Related Work Analysis}
Recent studies have investigated secure and reliable vehicular communication by adopting advanced trust management strategies. Quantum-inspired reinforcement learning, when integrated with permissioned blockchain systems, has shown promise in protecting decentralized vehicular networks while enabling adaptive resource allocation for vehicular services \cite{10949047, 10931853}. Variational quantum algorithms incorporating infeasibility constraints have been applied to improve collision avoidance routing in adversarial data environments, thus reducing vulnerabilities to packet manipulation and malicious path manipulation \cite{10715686}. Coevolutionary networking frameworks inspired by multiverse dynamics have enhanced the resilience of scale-free IoT topologies, which is crucial to maintaining stable V2X communication during link or node disruptions in dense urban settings \cite{10648662}. In parallel, decision-making architectures that utilize uncertainty quantification and distributional reinforcement learning have been employed to enhance the operational reliability of autonomous systems in scenarios such as uncontrolled intersections \cite{10155311}.  

Scalability in vehicular networks has intensified the challenges related to resource allocation and load balancing. To address these issues, edge–cloud continuum architectures have been developed to dynamically distribute computational workloads, supporting mobility prediction and adaptive traffic management in real time \cite{10103507, 10018121}. Digital twin-based Internet of Vehicles (IoV) systems have incorporated freshness-aware caching and quantum-inspired differential evolution to enhance task offloading, thereby reducing latency and energy consumption in highly dynamic network environments \cite{10530375}. Urban traffic optimization has also seen progress through the application of quantum annealing techniques to quadratic unconstrained binary optimization (QUBO) models, facilitating efficient traffic signal scheduling and routing strategies \cite{10239465}. Furthermore, attention-based forecasting models have improved the accuracy of spatiotemporal mobility predictions, encouraging proactive distribution of computational loads across edge and cloud nodes \cite{10768194}.  
The proposed Quantum-Inspired Vehicular Network Optimization Model (QIVNOM) addresses key limitations of prior approaches, particularly in terms of scalability, computational tractability, and adaptability in dynamic urban contexts. By incorporating probabilistic superposition and entanglement principles, QIVNOM enables parallelized optimization across both communication and decision-making layers, offering an integrated framework for vehicular network management. Tables~\ref{tab:comparative-analysis} present a comprehensive comparative analysis, highlighting how the proposed QIVNOM framework addresses critical limitations in existing approaches.

\renewcommand{\arraystretch}{1.8}
\begin{table}[!t]
\centering
\caption{Comparative Analysis of Existing Approaches and the Proposed QIVNOM Framework. Abbreviations: QSA = Quantum-Inspired Superposition Architecture, FTR = Fault-Tolerant Routing, MOO = Multi-Objective Optimization, FCC = Fog-Cloud Coordination, EAM = Energy-Aware Management, DTA = Dynamic Task Allocation, URT = Ultra-Low Response Time}
\label{tab:comparative-analysis}
\begin{tabular}{|p{1.2cm}|c|c|c|c|c|c|c|}
\hline
{Reference} & {QSA} & {FTR} & {MOO} & {FCC} & {EAM} & {DTA} & {URT} \\
\hline

\cite{10949047} & \ding{51} & \ding{55} & \ding{51} & \ding{55} & \ding{55} & \ding{51} & \ding{55} \\
\hline

\cite{10239465} & \ding{51} & \ding{55} & \ding{51} & \ding{55} & \ding{55} & \ding{55} & \ding{55} \\
\hline

\cite{10931853} & \ding{51} & \ding{51} & \ding{55} & \ding{55} & \ding{55} & \ding{55} & \ding{55} \\
\hline

\cite{10648662} & \ding{55} & \ding{51} & \ding{55} & \ding{55} & \ding{55} & \ding{55} & \ding{55} \\
\hline

\cite{10530375} & \ding{51} & \ding{55} & \ding{51} & \ding{51} & \ding{51} & \ding{51} & \ding{55} \\
\hline

\cite{10103507} & \ding{55} & \ding{55} & \ding{55} & \ding{51} & \ding{51} & \ding{51} & \ding{55} \\
\hline

\cite{10155311} & \ding{55} & \ding{51} & \ding{55} & \ding{55} & \ding{55} & \ding{55} & \ding{55} \\
\hline

{QIVNOM} & \ding{51} & \ding{51} & \ding{51} & \ding{51} & \ding{51} & \ding{51} & \ding{51} \\
\hline

\end{tabular}
\end{table}

\subsection{Problem Statement}

This research addresses the dual challenge of achieving ultra-low latency and fault-tolerant V2V/V2I communication, along with intelligent decision-making for urban traffic and energy management in smart cities \cite{10980011, 10288074}. The objective is to jointly optimize communication routing, resource distribution, and large-scale traffic control within a unified framework that meets the stringent real-time demands of critical vehicle safety applications \cite{10089194}. Existing methods often treat communication performance and traffic optimization as separate concerns, overlooking their interdependencies in a connected urban mobility system, where actions in one domain influence outcomes in the other \cite{10899852}. This gap underscores the need for an integrated solution that can coordinate vehicular communication and traffic management with high resilience and operational efficiency.

To address this, the proposed work introduces a computational framework inspired by quantum mechanics to explore the combinatorial solution space of urban-scale vehicular networks \cite{10841399}. The goal is to design a quantum-inspired algorithm that can simultaneously optimize interdependent variables such as data routing, traffic signal control, and energy-aware task assignment. The system is designed to function within a fog–cloud architecture, where latency-sensitive tasks are processed locally, and global coordination is maintained via the cloud infrastructure. This enables proactive fault mitigation, adaptation to real-time traffic dynamics, and continuity of service under high mobility and communication variability.

\subsection{Overview of the Proposed Approach}

The Quantum-Inspired Vehicular Network Optimization Model (QIVNOM) is structured as a multilayered system composed of edge-based vehicle nodes, fog computing intermediaries, and a centralized cloud coordinator. It integrates a quantum-inspired optimization algorithm that employs a probabilistic superposition of candidate solutions, iterative probability updates through interference principles, and entanglement modeling for correlated decision variables. This algorithm concurrently manages both network configurations and urban mobility control, achieving coordinated outcomes that consider latency, throughput, and energy constraints. The distributed architecture enables fog nodes to execute real-time, localized optimizations, while the cloud layer oversees the alignment of global strategies. QIVNOM is designed to efficiently explore high-dimensional decision spaces and converge toward near-optimal solutions within practical time bounds. The main contributions of this work are as follows:

\begin{itemize}
    \item Design of QIVNOM as an integrated framework for joint optimization of vehicular communication and urban decision making, operating in a distributed fog-cloud infrastructure.
    \item Development of a quantum-inspired algorithm based on superposition-driven probabilistic encoding and entangled variable modeling to enable efficient multi-objective optimization.
    \item Implementation of proactive redundancy and adaptive routing strategies to maintain continuous V2X communication and ensure system resilience under dynamic network and traffic conditions.
\end{itemize}

\subsection{Structure of the Article}
The remainder of this article is organized as follows. Section~\ref{sec:proposedapproach} details the architecture and components of the proposed Quantum-Inspired Vehicular Network Optimization Model. Section~\ref{sec:simulation} presents the simulation setup, the core evaluation parameters, and the comparative analysis against existing methods. Section~\ref{sec:ablation-analysis} provides an ablation study to examine the individual contributions of key system modules. Finally, Section~\ref{sec:conclusion} concludes the paper and outlines potential directions for future research.

\section{Proposed Approach: Quantum-Inspired Vehicular Network Optimization Model}
\label{sec:proposedapproach}
The proposed QIVNOM introduces a novel framework that integrates quantum-inspired optimization techniques with Vehicle-to-Everything (V2X) communication systems to improve the efficiency and resilience of smart city infrastructures. This section provides an in-depth overview of the multilayered architecture, comprising vehicle, fog, and cloud layers, along with the novel approach of co-optimizing latency, fault tolerance, and energy within V2X systems.

\subsection{Workflow of QIVNOM}

The QIVNOM workflow begins with the acquisition of real-time data from vehicles and nearby fog nodes. Vehicles transmit dynamic telemetry, including current location, velocity, and proximity to obstacles. Fog nodes aggregate this information and perform preliminary analysis to assess traffic conditions and identify local safety risks. This layer supports immediate responses when required, such as initiating re-routing or issuing alerts to nearby vehicles in the event of hazards.

\begin{figure*}[!t]
    \centering
    \includegraphics[width=0.90\linewidth]{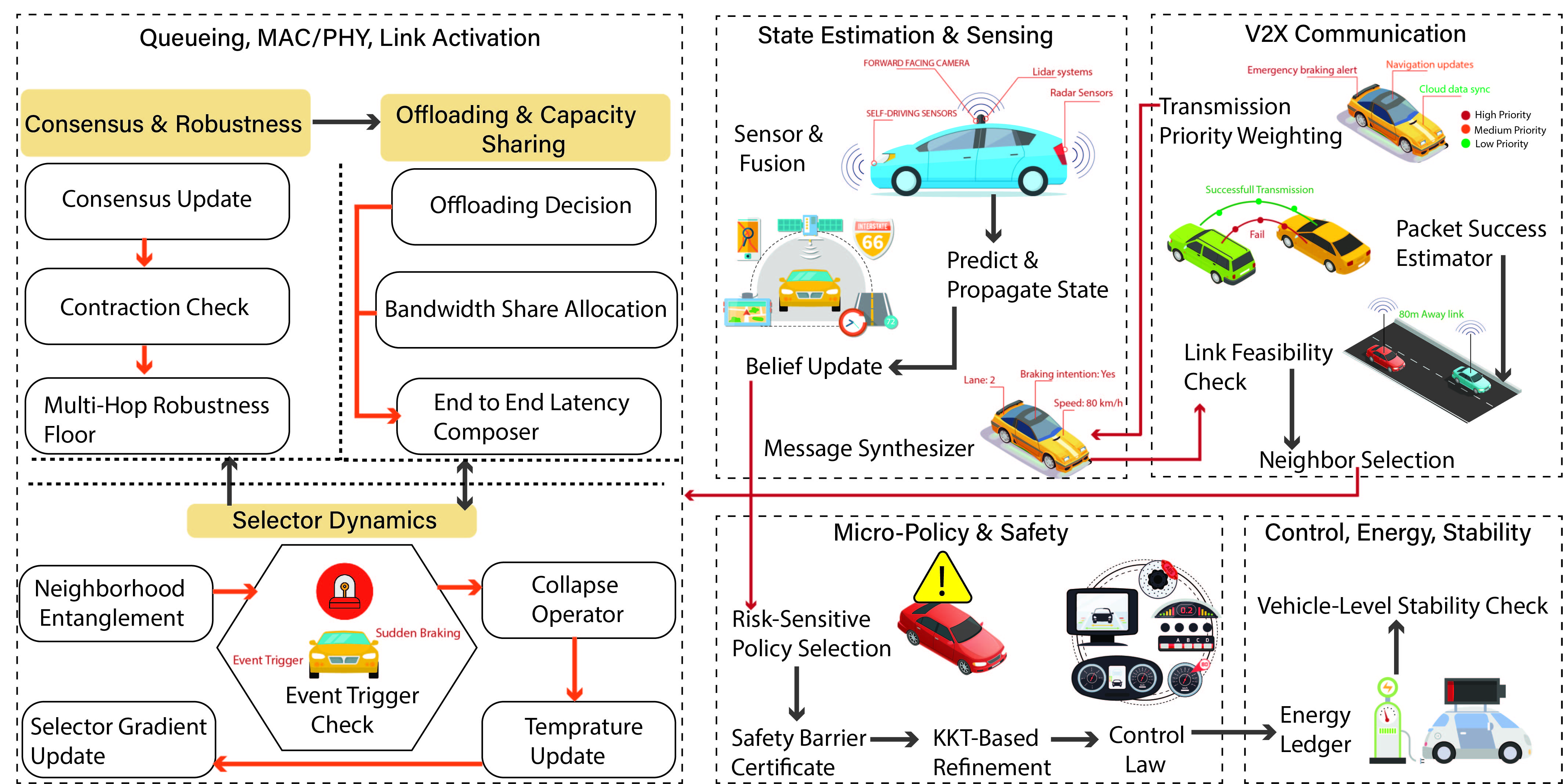}
    \caption{The Proposed QIVNOM Workflow.}
    \label{fig:placeholder}
\end{figure*}

Following data aggregation, the fog layer executes low-latency preprocessing and manages localized traffic events, such as congestion or collisions. Using quantum-inspired decision-making, the system optimizes regional parameters, such as signal timings and routing recommendations, through the use of probabilistic models. These models, guided by the quantum-inspired algorithm, support real-time decision-making to reduce latency and improve communication reliability. In the final stage, the cloud layer performs global optimization by integrating processed insights from fog nodes into a comprehensive city-wide model. The quantum-inspired algorithm evaluates multiple system objectives, including traffic flow, resource distribution, and energy utilization. The resulting optimized strategies are then communicated back to the edge and fog layers for implementation across vehicles and infrastructure components.

\subsection{Vehicle Layer: Edge Sensing and Communication}
The vehicle layer encodes edge sensing, V2V/V2I exchange, and micro-decisions as a constrained stochastic-control program with quantum-inspired local selectors. The algorithm~\ref{alg:vehicle-layer} operationalizes the edge pipeline by instantiating the sense-predict-update cycle through state propagation.

\begin{algorithm}[!t]
\caption{QIVNOM Vehicle-Layer Execution with Edge Sensing and Micro-Decisions}
\label{alg:vehicle-layer}
\SetKwInOut{Input}{Input}\SetKwInOut{Output}{Output}\SetKwInOut{Params}{Params}
\Input{Neighborhood $\mathcal{N}_i^{t}$, QoS targets, safety margins, and link metrics}
\Output{Control $u_i^{t}$, message set $\{m_{i\rightarrow j}^{t}\}$, offloading flag $o_i^{t}$}
\Params{Gains $K_i^{t},G_{ij}^{t}$, weights, deadlines, and budgets}

\For{$t=0,1,\dots,T-1$}{
  Obtain measurement via \ref{eq:veh_obs}; predict and propagate the state using \ref{eq:veh_state}; refine belief using \ref{eq:veh_belief_update}\;

  Form sparse message $m_{i\rightarrow j}^{t}$ using \ref{eq:veh_msg}\;
  \eIf{latency constraint \ref{eq:veh_latency} holds {and} link indicator \ref{eq:veh_graph} is true}{
    transmit to neighbors with success probability from \ref{eq:veh_linkprob}\;
  }{
    select alternative neighbor with highest \ref{eq:veh_linkprob} and defer transmission\;
  }

  Update local queue using \ref{eq:veh_queue}; choose PHY/MAC profile via \ref{eq:veh_sched}\;
  Compute activation probabilities $\theta_{ij}^{t}$ using \ref{eq:veh_activation}\;

  Determine micro-action by CVaR policy \ref{eq:veh_policy_cvar} subject to safety certificate \ref{eq:veh_safety_cbf}\;
  Refine action via KKT stationarity \ref{eq:veh_kkt}\;

  Update energy ledger with \ref{eq:veh_energy}\;
  Apply control and message feedback using \ref{eq:veh_control}\;

  Update local selector $\psi_i^{t}$ with \ref{eq:veh_psi_update}; inject neighborhood influence using \ref{eq:veh_entangle}\;
  \If{trigger condition \ref{eq:veh_trigger} holds}{
     collapse to $e_{k^\star}$ using \ref{eq:veh_collapse}; update temperature via \ref{eq:veh_temp}\;
  }

  Decide offloading $o_i^{t}$ via \ref{eq:veh_offload_decision}; compute end-to-end latency using \ref{eq:veh_offload_latency}\;
  Allocate bandwidth share and compute link rate using \ref{eq:veh_bandwidth}\;

  Set transmission priority weight via \ref{eq:veh_priority}; estimate packet success probability via \ref{eq:veh_packetloss}\;

  Update consensus state using \ref{eq:veh_consensus}; verify contraction using \ref{eq:veh_convergence}\;
  Ensure path robustness floor via \ref{eq:veh_robust_path}\;
  Check Lyapunov decrement using \ref{eq:veh_lyapunov}; \If{violation detected}{tighten safety margin in \ref{eq:veh_safety_cbf} and reduce step in \ref{eq:veh_psi_update}}
}
\end{algorithm}
\begin{subequations}
\renewcommand{\theequation}{\theparentequation\alph{equation}}
\begin{equation}
\label{eq:veh_state}
x_i^{t+1}=\Phi_i^{t}x_i^{t}+\Gamma_i^{t}u_i^{t}+\sum_{j\in\mathcal{N}_i^{t}}B_{ij}^{t}m_{j\rightarrow i}^{t}+w_i^{t}
\end{equation}
\begin{equation}
\label{eq:veh_obs}
y_i^{t}=H_i^{t}x_i^{t}+n_i^{t}
\end{equation}
\begin{equation}
\label{eq:veh_belief_update}
b_i^{t+}\propto b_i^{t-}\exp\!\Big(-\tfrac{1}{2}\big\|y_i^{t}-H_i^{t}\hat{x}_i^{t}\big\|_{(R_i^{t})^{-1}}^{2}\Big)
\end{equation} \end{subequations}
In \ref{eq:veh_state}, $x_i^{t}$ is the kinematic augmented state of vehicle $i$, $u_i^{t}$ is the low level control, $m_{j\rightarrow i}^{t}$ is the message received from neighbor $j$, and $w_i^{t}$ is the noise of the bounded process. In \ref{eq:veh_obs}, $y_i^{t}$ is the fused measurement, $H_i^{t}$ the sensing map, and $n_i^{t}$ measurement noise. In \ref{eq:veh_belief_update}, $b_i^{t\pm}$ denote prior/posterior beliefs, $\hat{x}_i^{t}$ is the predictor, and $R_i^{t}$ is the innovation covariance. Next, we impose message sparsity \ref{eq:veh_msg}, latency feasibility \ref{eq:veh_latency}, link existence \ref{eq:veh_graph}, and probabilistic link use \ref{eq:veh_linkprob}.
\begin{subequations}
\renewcommand{\theequation}{\theparentequation\alph{equation}}
\begin{equation}
\label{eq:veh_msg}
m_{i\rightarrow j}^{t}=\Pi_{ij}^{t}S_i^{t}\hat{x}_i^{t},\qquad \|m_{i\rightarrow j}^{t}\|_{0}\le C_{ij}^{t}
\end{equation}
\begin{equation}
\label{eq:veh_latency}
\tau_{ij}^{t}=\frac{\ell_{ij}^{t}}{r_{ij}^{t}}+\frac{q_{j}^{t}}{\mu_{j}^{t}}\le \bar{\tau}_{ij}
\end{equation}
\begin{equation}
\label{eq:veh_graph}
A_{ij}^{t}=\mathbf{1}\{\mathrm{SNR}_{ij}^{t}\ge\gamma_{\mathrm{th}}\ \wedge\ d_{ij}^{t}\le d_{\max}\}
\end{equation}
\begin{equation}
\label{eq:veh_linkprob}
\pi_{ij}^{t}=\Pr\{A_{ij}^{t}=1\mid \mathrm{SNR}_{ij}^{t},d_{ij}^{t}\}
\end{equation} \end{subequations}
In \ref{eq:veh_msg}, $S_i^{t}$ compresses features, $\Pi_{ij}^{t}$ masks fields, and $C_{ij}^{t}$ bounds sparsity. In \ref{eq:veh_latency}, $\ell_{ij}^{t}$ is payload length, $r_{ij}^{t}$ the physical rate, $q_j^{t}$ the service queue, $\mu_j^{t}$ the service rate, $\bar{\tau}_{ij}$ the deadline. In \ref{eq:veh_graph}, $A_{ij}^{t}$ flags usable links via SNR threshold $\gamma_{\mathrm{th}}$ and range $d_{\max}$. In \ref{eq:veh_linkprob}, $\pi_{ij}^{t}$ quantifies link reliability. Local queue evolution \ref{eq:veh_queue}, schedule selection \ref{eq:veh_sched}, quantum-inspired activation \ref{eq:veh_activation}, risk-aware micro-policy \ref{eq:veh_policy_cvar}, and safety certificates \ref{eq:veh_safety_cbf} govern fast responses.
\begin{subequations}
\renewcommand{\theequation}{\theparentequation\alph{equation}}
\begin{equation}
\label{eq:veh_queue}
q_{i}^{t+1}=\max\{0,q_{i}^{t}-\mu_{i}^{t}\}+\sum_{j}\lambda_{ji}^{t}
\end{equation}
\begin{equation}
\label{eq:veh_sched}
r_{ij}^{t}=\sum_{\ell}\sigma_{i\ell}^{t}R_{ij}^{(\ell),t},\quad \sigma_{i\ell}^{t}\in\{0,1\},\ \sum_{\ell}\sigma_{i\ell}^{t}\le 1
\end{equation}
\begin{equation}
\label{eq:veh_activation}
\theta_{ij}^{t}=|\psi_{i}^{t}(j)|^{2},\qquad \sum_{j}\theta_{ij}^{t}=1
\end{equation}
\begin{equation}
\label{eq:veh_policy_cvar}
\pi_{i}^{t}=\arg\min_{a\in\mathcal{A}_{i}}\ \mathrm{CVaR}_{\alpha}\!\big(J_{i}(x_{i}^{t},a,\xi^{t})\big)
\end{equation}
\begin{equation}
\label{eq:veh_safety_cbf}
\dot{h}_{i}(x_{i}^{t},u_{i}^{t})+\kappa_{i}\,h_{i}(x_{i}^{t})\ge 0
\end{equation} \end{subequations}
In \ref{eq:veh_queue}, $\lambda_{ji}^{t}$ is the arrival rate. In \ref{eq:veh_sched}, $\sigma_{i\ell}^{t}$ chooses one PHY/MAC profile $\ell$ with achievable $R_{ij}^{(\ell),t}$. Equation \ref{eq:veh_activation} maps the superposition vector $\psi_i^{t}$ to link-use probabilities $\theta_{ij}^{t}$. In \ref{eq:veh_policy_cvar}, $J_i$ is the episodic cost and $\xi^{t}$ exogenous randomness. In \ref{eq:veh_safety_cbf}, $h_i$ enforces forward invariance with gain $\kappa_i$. Optimality, energy accounting, and wavefunction transport appear in \ref{eq:veh_kkt}–\ref{eq:veh_psi_update}.
\begin{subequations}
\renewcommand{\theequation}{\theparentequation\alph{equation}}
\begin{equation}
\label{eq:veh_kkt}
\nabla_{a}\mathcal{L}_{i}(a,\lambda)=\nabla_{a}\hat{J}_{i}(a)+\lambda\nabla_{a}g_{i}(a)=0,\quad \lambda g_{i}(a)=0,\ \lambda\ge 0
\end{equation}
\begin{equation}
\label{eq:veh_energy}
e_{i}^{t+1}=e_{i}^{t}+\chi_{\mathrm{drive}}\|u_{i}^{t}\|_{2}^{2}+\chi_{\mathrm{comm}}\sum_{j}r_{ij}^{t}
\end{equation}
\begin{equation}
\label{eq:veh_psi_update}
\psi_{i}^{t+1}=\frac{M_{i}^{t}\psi_{i}^{t}}{\|M_{i}^{t}\psi_{i}^{t}\|_{2}},\qquad M_{i}^{t}=\exp\!\big(-\eta_{t}\nabla_{\psi_{i}}\tilde{J}_{i}(\psi_{i}^{t})\big)
\end{equation} \end{subequations}
In \ref{eq:veh_kkt}, $\hat{J}_i$ is the surrogate stage cost and $g_i$ constraints with multiplier $\lambda$. In \ref{eq:veh_energy}, $e_i^{t}$ is the energy ledger, with drive and communication coefficients $\chi_{\mathrm{drive}},\chi_{\mathrm{comm}}$. In \ref{eq:veh_psi_update}, $M_i^{t}$ is the update operator with stepsize $\eta_t$. Neighborhood entanglement \ref{eq:veh_entangle}, collapse trigger \ref{eq:veh_collapse}, adaptive temperature \ref{eq:veh_temp}, offloading rule \ref{eq:veh_offload_decision}, latency composition \ref{eq:veh_offload_latency}, and capacity sharing \ref{eq:veh_bandwidth} bind sensing to computation.
\begin{subequations}
\renewcommand{\theequation}{\theparentequation\alph{equation}}
\begin{equation}
\label{eq:veh_entangle}
\psi_{i}^{t+\frac{1}{2}}=\psi_{i}^{t}\odot\!\!\!\prod_{j\in\mathcal{N}_{i}^{t}}\!\!\Big(I+\mathcal{E}_{ij}^{t}\,\mathrm{diag}(\psi_{j}^{t})\Big)
\end{equation}
\begin{equation}
\label{eq:veh_collapse}
\psi_{i}^{t+}=e_{k^{\star}},\quad k^{\star}=\arg\max_{k}\psi_{i,k}^{t+\frac{1}{2}}\ \text{ when }\ J_{i}^{t}-J_{i}^{t-1}\ge \eta_{i}^{t}
\end{equation}
\begin{equation}
\label{eq:veh_temp}
T_{i}^{t+1}=\beta T_{i}^{t}+(1-\beta)\frac{\mathrm{var}(J_{i}^{t-\tau:t})}{1+\mathrm{var}(J_{i}^{t-\tau:t})}
\end{equation}
\begin{equation}
\label{eq:veh_offload_decision}
o_{i}^{t}=\mathbf{1}\{L_{\mathrm{loc},i}^{t}-L_{\mathrm{fog},i}^{t}\ge \delta_{i}^{t}\}
\end{equation}
\begin{equation}
\label{eq:veh_offload_latency}
L_{i}^{t}=(1-o_{i}^{t})L_{\mathrm{loc},i}^{t}+o_{i}^{t}\big(L_{\mathrm{upl},i}^{t}+L_{\mathrm{proc},i}^{t}+L_{\mathrm{down},i}^{t}\big)
\end{equation}
\begin{equation}
\label{eq:veh_bandwidth}
r_{ij}^{t}=\alpha_{i}^{t}B\log_{2}\big(1+\mathrm{SINR}_{ij}^{t}\big),\qquad \sum_{i}\alpha_{i}^{t}\le 1
\end{equation} \end{subequations}
In \ref{eq:veh_entangle}, $\mathcal{E}_{ij}^{t}$ injects neighbor influence. In \ref{eq:veh_collapse}, $e_{k^{\star}}$ is the one-hot selector when cost jump exceeds $\eta_{i}^{t}$. In \ref{eq:veh_temp}, $T_{i}^{t}$ adapts to cost volatility with smoothing $\beta$. In \ref{eq:veh_offload_decision}–\ref{eq:veh_offload_latency}, $o_i^{t}$ decides fog offload using latency components. In \ref{eq:veh_bandwidth}, $B$ is bandwidth and $\alpha_{i}^{t}$ the share. Priority and packet success are encoded by \ref{eq:veh_priority} and \ref{eq:veh_packetloss}.
\begin{subequations}
\renewcommand{\theequation}{\theparentequation\alph{equation}}
\begin{equation}
\label{eq:veh_priority}
\omega_{i}^{t}=\alpha_{s}s_{i}^{t}+\alpha_{f}f_{i}^{t}+\alpha_{\mathrm{st}}\mathrm{stale}_{i}^{t},\qquad \sum_{i}\omega_{i}^{t}=1
\end{equation}
\begin{equation}
\label{eq:veh_packetloss}
p_{\mathrm{succ},ij}^{t}=\Big(1+\exp\{-\xi(\gamma_{ij}^{t}-\gamma_{0})\}\Big)^{-1}\gamma_{\mathrm{code}}
\end{equation} \end{subequations}
In \ref{eq:veh_priority}, $s_i^{t}$ is safety urgency, $f_i^{t}$ fault risk, and $\mathrm{stale}_{i}^{t}$ data staleness. In \ref{eq:veh_packetloss}, $\gamma_{ij}^{t}$ is SNR, $\gamma_{0}$ threshold, $\xi$ steepness, and $\gamma_{\mathrm{code}}$ coding gain. Consensus \ref{eq:veh_consensus}, contraction \ref{eq:veh_convergence}, path robustness \ref{eq:veh_robust_path}, and event trigger \ref{eq:veh_trigger} coordinate group behavior.
\begin{subequations}
\renewcommand{\theequation}{\theparentequation\alph{equation}}
\begin{equation}
\label{eq:veh_consensus}
\xi_{i}^{t+1}=\xi_{i}^{t}+\sum_{j\in\mathcal{N}_{i}^{t}}w_{ij}^{t}\big(\xi_{j}^{t}-\xi_{i}^{t}\big)+\eta_{i}^{t}\big(\hat{\xi}_{i}^{t}-\xi_{i}^{t}\big)
\end{equation}
\begin{equation}
\label{eq:veh_convergence}
\sum_{i}\|\xi_{i}^{t+1}-\bar{\xi}^{t+1}\|_{2}^{2}\le \rho \sum_{i}\|\xi_{i}^{t}-\bar{\xi}^{t}\|_{2}^{2}
\end{equation}
\begin{equation}
\label{eq:veh_robust_path}
\rho_{i}^{t}=\min_{p\in\mathcal{P}_{i}^{t}}\ \prod_{(a,b)\in p}\theta_{ab}^{t},\qquad \rho_{i}^{t}\ge \rho_{\min}
\end{equation}
\begin{equation}
\label{eq:veh_trigger}
\Delta J_{i}^{t}=J_{i}^{t}-J_{i}^{t-1}\ge \eta_{i}^{t}\ \Rightarrow\ \text{invoke } \ref{eq:veh_collapse}
\end{equation} \end{subequations}
In \ref{eq:veh_consensus}, $w_{ij}^{t}$ are Metropolis weights and $\hat{\xi}_{i}^{t}$ is a local innovation. In \ref{eq:veh_convergence}, $\rho\!<\!1$ sets contraction. In \ref{eq:veh_robust_path}, $\rho_{i}^{t}$ is minimum multi-hop success with floor $\rho_{\min}$. In \ref{eq:veh_trigger}, cost spikes activate the collapse operator in \ref{eq:veh_collapse}. Finally, the control map \ref{eq:veh_control} and Lyapunov decrement \ref{eq:veh_lyapunov} close the loop.
\begin{subequations}
\renewcommand{\theequation}{\theparentequation\alph{equation}}
\begin{equation}
\label{eq:veh_control}
u_{i}^{t}=K_{i}^{t}x_{i}^{t}+\sum_{j\in\mathcal{N}_{i}^{t}}G_{ij}^{t}m_{j\rightarrow i}^{t},\qquad \|u_{i}^{t}\|_{\infty}\le u_{\max}
\end{equation}
\begin{equation}
\label{eq:veh_lyapunov}
V_{i}^{t+1}-V_{i}^{t}\le -\lambda_{i}\|x_{i}^{t}\|_{2}^{2}+\chi\|w_{i}^{t}\|_{2}^{2}
\end{equation} \end{subequations}
In \ref{eq:veh_control}, $K_{i}^{t}$ and $G_{ij}^{t}$ are state and message gains with bounded actuation $u_{\max}$. In \ref{eq:veh_lyapunov}, $V_i^{t}$ decreases by margin $\lambda_i$ under disturbance weight $\chi$, ensuring vehicle-level stability under the full communication–control stack.
\subsection{Fog Layer: Edge Computing and Localized Optimization}

Fog nodes condense heterogeneous vehicular data streams, execute short-horizon optimizations, and provide early responses within constrained latency and energy budgets. Aggregation, sketching, privacy preservation, hazard scoring, rerouting, and dynamic compute–storage assignment are formalized as coupled stochastic programs that interface directly with vehicle-level decisions and upstream cloud coordination. Sensed signals are aggregated in \ref{eq:fog_agg}, a nonlinear sketch is constructed in \ref{eq:fog_sketch}, and calibrated noise is introduced in \ref{eq:fog_privacy} to ensure privacy compliance.
\begin{subequations}
\renewcommand{\theequation}{\theparentequation\alph{equation}}
\begin{equation}
\label{eq:fog_agg}
z_{f}^{t}=\sum_{i\in\mathcal{N}_{f}^{t}}W_{fi}^{t}y_{i}^{t}+U_{f}^{t}\bar{y}_{f}^{t}
\end{equation}
\begin{equation}
\label{eq:fog_sketch}
s_{f}^{t}=P_{f}^{t}\tanh\!\big(\Omega_{f}^{t}z_{f}^{t}+b_{f}^{t}\big)
\end{equation}
\begin{equation}
\label{eq:fog_privacy}
\tilde{s}_{f}^{t}=s_{f}^{t}+\sigma_{f}^{t}\varepsilon_{f}^{t},\qquad \varepsilon_{f}^{t}\sim\mathcal{N}(0,I)
\end{equation} \end{subequations}
In \ref{eq:fog_agg}, $z_{f}^{t}$ denotes the fused vector at fog node $f$, where $W_{fi}^{t}$ represents the adaptive weights assigned to vehicle measurements $y_{i}^{t}$, and $U_{f}^{t}\bar{y}_{f}^{t}$ encodes inputs from local roadside sensing infrastructure. In \ref{eq:fog_sketch}, $P_{f}^{t}$ and $\Omega_{f}^{t}$ are projection operators, $b_{f}^{t}$ is a bias term, and $\tanh(\cdot)$ is applied to stabilize extreme values. In \ref{eq:fog_privacy}, $\sigma_{f}^{t}$ defines the privacy budget, and $\varepsilon_{f}^{t}$ denotes zero-mean Gaussian noise added for differential privacy. Hazard intensity is computed in \ref{eq:fog_hazard}, candidate detours are scored in \ref{eq:fog_route_score}, the selected route is determined in \ref{eq:fog_route_pick}, and subchannel scheduling is addressed in \ref{eq:fog_sched_subch}.
\begin{subequations}
\renewcommand{\theequation}{\theparentequation\alph{equation}}
\begin{equation}
\label{eq:fog_hazard}
r_{fi}^{t}=\ln\!\big(1+\exp(\alpha_{f}^{t}\|D_{fi}^{t}z_{f}^{t}\|_{1}+\beta_{f}^{t})\big)
\end{equation}
\begin{equation}
\label{eq:fog_route_score}
\rho_{i\ell}^{t}=w_{1}^{t}\hat{T}_{i\ell}^{t}+w_{2}^{t}\hat{c}_{i\ell}^{t}+w_{3}^{t}\big(\hat{b}_{i\ell}^{t}\big)^{-1}
\end{equation}
\begin{equation}
\label{eq:fog_route_pick}
a_{i}^{t}=\arg\min_{\ell\in\mathcal{L}_{i}^{t}}\ \rho_{i\ell}^{t}\quad \text{s.t.}\quad r_{fi}^{t}\le \tau_{f}^{t}
\end{equation}
\begin{equation}
\label{eq:fog_sched_subch}
R_{fi\rightarrow j}^{t}=\sum_{u}\sigma_{fu}^{t}\, \mathcal{R}_{fi\rightarrow j}^{(u),t},\qquad \sum_{u}\sigma_{fu}^{t}\le 1,\ \sigma_{fu}^{t}\in\{0,1\}
\end{equation} \end{subequations}
In \ref{eq:fog_hazard}, $r_{fi}^{t}$ is a smooth hazard score, $\alpha_{f}^{t}$ and $\beta_{f}^{t}$ tune sensitivity, and $D_{fi}^{t}$ extracts salient features. In \ref{eq:fog_route_score}, $\hat{T}_{i\ell}^{t}$ is predicted travel time, $\hat{c}_{i\ell}^{t}$ congestion, $\hat{b}_{i\ell}^{t}$ network bandwidth; $w_{1}^{t},w_{2}^{t},w_{3}^{t}$ are normalized weights. In \ref{eq:fog_route_pick}, $a_{i}^{t}$ picks a feasible detour under threshold $\tau_{f}^{t}$. In \ref{eq:fog_sched_subch}, $\sigma_{fu}^{t}$ activates at most one subchannel $u$ with achievable rate $\mathcal{R}_{fi\rightarrow j}^{(u),t}$. Dynamic compute–storage allocation is posed via capacity-aware shares in \ref{eq:fog_cpu}, cache fractions in \ref{eq:fog_cache}, a local objective in \ref{eq:fog_obj}, dual updates in \ref{eq:fog_dual}, and a proximal step in \ref{eq:fog_prox}.
\begin{subequations}
\renewcommand{\theequation}{\theparentequation\alph{equation}}
\begin{equation}
\label{eq:fog_cpu}
\sum_{k\in\mathcal{T}_{f}^{t}}\pi_{fk}^{t}\le C_{f}^{\mathrm{cpu}},\qquad \pi_{fk}^{t}\ge 0
\end{equation}
\begin{equation}
\label{eq:fog_cache}
\sum_{k\in\mathcal{T}_{f}^{t}}\kappa_{fk}^{t}\le C_{f}^{\mathrm{mem}},\qquad 0\le \kappa_{fk}^{t}\le 1
\end{equation}
\begin{equation}
\label{eq:fog_obj}
\mathcal{J}_{f}^{t}=\sum_{k\in\mathcal{T}_{f}^{t}} \omega_{k}^{t}\Big(\alpha L_{fk}^{t}+\beta E_{fk}^{t}+\gamma S_{fk}^{t}\Big)
\end{equation}
\begin{equation}
\label{eq:fog_dual}
\lambda_{f}^{t+1}=\big[\lambda_{f}^{t}+\eta_{f}^{t}\big(\sum_{k}\pi_{fk}^{t}-C_{f}^{\mathrm{cpu}}\big)\big]_{+}
\end{equation}
\begin{equation}
\label{eq:fog_prox}
\Pi_{f}^{t+1}=\operatorname{prox}_{\eta_{f}^{t}\mathcal{R}_{f}}\!\Big(\Pi_{f}^{t}-\eta_{f}^{t}\nabla_{\Pi}\mathcal{L}_{f}^{t}\Big)
\end{equation} \end{subequations}
In \ref{eq:fog_cpu}, $\pi_{fk}^{t}$ is CPU share for task $k$, bounded by capacity $C_{f}^{\mathrm{cpu}}$. In \ref{eq:fog_cache}, $\kappa_{fk}^{t}$ is storage fraction with memory cap $C_{f}^{\mathrm{mem}}$. In \ref{eq:fog_obj}, $L_{fk}^{t}$ is latency, $E_{fk}^{t}$ energy, and $S_{fk}^{t}$ storage cost with positive weights $\alpha,\beta,\gamma$ and task priority $\omega_{k}^{t}$. In \ref{eq:fog_dual}, $\lambda_{f}^{t}$ is a nonnegative dual for CPU feasibility with step $\eta_{f}^{t}$. In \ref{eq:fog_prox}, $\Pi_{f}^{t}$ stacks decision variables, $\mathcal{R}_{f}$ is a convex regularizer, and $\mathcal{L}_{f}^{t}$ the augmented objective. Queueing, delay bounds, cache reuse, energy budget, stability, and alert triggering are described by \ref{eq:fog_queue}, \ref{eq:fog_delay}, \ref{eq:fog_hit}, \ref{eq:fog_energy}, \ref{eq:fog_lyap}, and \ref{eq:fog_alert}.
\begin{subequations}
\renewcommand{\theequation}{\theparentequation\alph{equation}}
\begin{equation}
\label{eq:fog_queue}
Q_{f}^{t+1}=\max\{0,Q_{f}^{t}+A_{f}^{t}-S_{f}^{t}\}
\end{equation}
\begin{equation}
\label{eq:fog_delay}
L_{fi}^{t}=\frac{Z_{fi}^{t}}{\mu_{f}^{t}-\Lambda_{fi}^{t}}+\Delta_{fi}^{t},\qquad \mu_{f}^{t}>\Lambda_{fi}^{t}
\end{equation}
\begin{equation}
\label{eq:fog_hit}
h_{fk}^{t}=1-\exp\!\big(-\nu_{f}^{t}\,\mathrm{req}_{fk}^{t-\tau:t}\big)
\end{equation}
\begin{equation}
\label{eq:fog_energy}
E_{f}^{t+1}=E_{f}^{t}+\eta_{\mathrm{cpu}}\sum_{k}\pi_{fk}^{t}+\eta_{\mathrm{tx}}\sum_{i,j}R_{fi\rightarrow j}^{t}
\end{equation}
\begin{equation}
\label{eq:fog_lyap}
V_{f}^{t+1}-V_{f}^{t}\le -\lambda_{f}\|Q_{f}^{t}\|_{2}^{2}+\chi_{f}\|A_{f}^{t}\|_{2}^{2}
\end{equation}
\begin{equation}
\label{eq:fog_alert}
\zeta_{fi}^{t}=\mathbf{1}\{r_{fi}^{t}\ge \tau_{f}^{t}\}
\end{equation} \end{subequations}
In \ref{eq:fog_queue}, $Q_{f}^{t}$ is the backlog, $A_{f}^{t}$ arrivals, and $S_{f}^{t}$ service. In \ref{eq:fog_delay}, $L_{fi}^{t}$ is local delay for vehicle $i$, with service rate $\mu_{f}^{t}$, load $\Lambda_{fi}^{t}$, residual term $Z_{fi}^{t}$, and processing overhead $\Delta_{fi}^{t}$. In \ref{eq:fog_hit}, $h_{fk}^{t}$ is cache hit probability given request intensity $\mathrm{req}_{fk}^{t-\tau:t}$ and reuse factor $\nu_{f}^{t}$. In \ref{eq:fog_energy}, $E_{f}^{t}$ tracks energy; $\eta_{\mathrm{cpu}}$ and $\eta_{\mathrm{tx}}$ weight compute and transmission. In \ref{eq:fog_lyap}, $V_{f}^{t}$ is a Lyapunov function with margin $\lambda_{f}$ and disturbance weight $\chi_{f}$. In \ref{eq:fog_alert}, $\zeta_{fi}^{t}$ triggers local broadcasts when the score $r_{fi}^{t}$ exceeds the threshold.

\noindent Equations \ref{eq:fog_agg}–\ref{eq:fog_privacy} feed hazard scoring \ref{eq:fog_hazard}; the routing choice \ref{eq:fog_route_pick} uses the score from \ref{eq:fog_hazard} and the metric in \ref{eq:fog_route_score}; scheduled rates in \ref{eq:fog_sched_subch} influence energy in \ref{eq:fog_energy} and backlog in \ref{eq:fog_queue}; allocations \ref{eq:fog_cpu}–\ref{eq:fog_prox} control delay \ref{eq:fog_delay}, cache reuse \ref{eq:fog_hit}, and stability \ref{eq:fog_lyap}; alerts in \ref{eq:fog_alert} close the local response loop.

\subsection{Cloud Layer: Global Coordination and Heavy Analytics}
The cloud layer consolidates summarized fog streams, learns long-horizon dynamics, and coordinates city-wide actions through quantum-inspired parallel optimization with constraint-aware resource policies. Algorithm~\ref{alg:cloud-layer} operationalizes cloud-level aggregation, long-horizon forecasting, multi-objective control, quantum-inspired population search, and entropic transport assignment, issuing deployments only when chance-constrained reliability and Lyapunov stability conditions are satisfied.

\begin{algorithm}[t]
\caption{QIVNOM Cloud-Layer Global Coordination and Heavy Analytics}
\label{alg:cloud-layer}
\SetKwInOut{Input}{Input}\SetKwInOut{Output}{Output}\SetKwInOut{Params}{Params}
\Input{Fog summaries $\{\tilde{s}_{f}^{t}\}_{f\in \mathcal{F}_{t}}$, risk level $\delta$, horizons $H,W$, marginals $(\mu,\nu)$}
\Output{Global plan index $k^{\star}$, assignment $\Pi^{\star}$, updated $(X^{t},\Theta^{t})$}
\Params{Stepsizes $(\eta_{x},\eta_{\lambda},\eta_{t},\eta_{p})$, smoothing $\beta,\rho$, penalties $(\lambda,\varepsilon)$}

\For{$t=0,1,\dots,T-1$}{
  Compute $Z^{t}$ using \ref{eq:cloud_agg}; compute $\Xi^{t}$ using \ref{eq:cloud_lift}; compute confidences $\{\omega_{f}^{t}\}$ using \ref{eq:cloud_conf}\;
  \ForEach{$f\in \mathcal{F}_{t}$}{reweight $\tilde{s}_{f}^{t}\leftarrow \omega_{f}^{t}\tilde{s}_{f}^{t}$}

  Update $X^{t+1}$ using \ref{eq:cloud_state}; compute observation $Y^{t}$ and residual with \ref{eq:cloud_obs}\;

  Update predictor $K^{t+1}$ using \ref{eq:cloud_koopman}\;
  Form loss $\mathcal{J}^{t}(\Theta)$ using \ref{eq:cloud_loss}; update $\Theta^{t+1}$ using \ref{eq:cloud_theta}; update second moment $M^{t+1}$ using \ref{eq:cloud_moment}\;

  Compute $J_{\mathrm{city}}^{t}$ using \ref{eq:cloud_mobj}; build $\mathcal{L}(\cdot)$ using \ref{eq:cloud_lagrangian}; update $(X^{t+1},\lambda^{t+1})$ using \ref{eq:cloud_pd}\;

  Update population $P^{t+1}$ using \ref{eq:cloud_qpop}; update temperature $T^{t+1}$ using \ref{eq:cloud_temp}\;
  Form soft plan $\{\pi_{k}^{t}\}$ and pick $k^{\star}$ using \ref{eq:cloud_sample}\;

  Solve $\Pi^{\star}$ using \ref{eq:cloud_ot}\;

  \eIf{chance constraint \ref{eq:cloud_cc} holds {and} Lyapunov condition \ref{eq:cloud_lyap} holds}{
     dispatch plan $k^{\star}$ with assignment $\Pi^{\star}$; continue\;
  }{
     \tcp{Adaptive repair: adjust trade-offs and search temperature}
     Update weights $w_{q}^{t}\leftarrow \mathrm{adjust}(w_{q}^{t})$; tighten risk level $\delta\leftarrow \mathrm{shrink}(\delta)$; modify $\eta_{p},\varepsilon$ and repeat population and OT steps\;
  }
}
\end{algorithm}

Global aggregation, forecasting, multi-objective control, population-based exploration, and assignment to fog nodes are formalized, incorporating explicit stability and reliability surrogates. Fog-level summaries are fused in \ref{eq:cloud_agg}, feature representations are lifted in \ref{eq:cloud_lift}, and source confidence is scored in \ref{eq:cloud_conf}.
\begin{subequations}
\renewcommand{\theequation}{\theparentequation\alph{equation}}
\begin{equation}
\label{eq:cloud_agg}
Z^{t}=\sum_{f\in\mathcal{F}_{t}}A_{f}^{t}\,\tilde{s}_{f}^{t}
\end{equation}
\begin{equation}
\label{eq:cloud_lift}
\Xi^{t}=\Phi^{t}\,\sigma\!\big(G^{t}Z^{t}+b^{t}\big)
\end{equation}
\begin{equation}
\label{eq:cloud_conf}
\omega_{f}^{t}=\frac{\exp\!\big(-\kappa_{1}\,\mathrm{var}(\tilde{s}_{f}^{t-\tau:t})-\kappa_{2}\,L_{f}^{t}\big)}{\sum_{g}\exp\!\big(-\kappa_{1}\,\mathrm{var}(\tilde{s}_{g}^{t-\tau:t})-\kappa_{2}\,L_{g}^{t}\big)}
\end{equation} \end{subequations}
In \ref{eq:cloud_agg}, $A_{f}^{t}$ integrates the fog summary $\tilde{s}_{f}^{t}$ into the global representation $Z^{t}$. In \ref{eq:cloud_lift}, $\sigma(\cdot)$ denotes a bounded activation function, with $\Phi^{t}$, $G^{t}$, and $b^{t}$ as learned parameters. In \ref{eq:cloud_conf}, $\omega_{f}^{t}$ assigns a weight to fog node $f$ based on its variance window and latency $L_{f}^{t}$. A city-level state model is adopted in \ref{eq:cloud_state}, with corresponding observations defined in \ref{eq:cloud_obs}.
\begin{subequations}
\renewcommand{\theequation}{\theparentequation\alph{equation}}
\begin{equation}
\label{eq:cloud_state}
X^{t+1}=F^{t}X^{t}+U^{t}+\sum_{f}H_{f}^{t}\,\tilde{s}_{f}^{t}+\varepsilon^{t}
\end{equation}
\begin{equation}
\label{eq:cloud_obs}
Y^{t}=C^{t}X^{t}+v^{t}
\end{equation} \end{subequations}
In \ref{eq:cloud_state}, $X^{t}$ represents the global system state, $U^{t}$ serves as the control proxy, $H_{f}^{t}$ incorporates signals from fog nodes, and $\varepsilon^{t}$ denotes the process noise. In \ref{eq:cloud_obs}, $Y^{t}$ is the observable output, perturbed by measurement noise $v^{t}$. A lifted predictor is estimated in \ref{eq:cloud_koopman}, a horizon-based loss is defined in \ref{eq:cloud_loss}, parameters are updated in \ref{eq:cloud_theta}, and second moments are adapted in \ref{eq:cloud_moment}.
\begin{subequations}
\renewcommand{\theequation}{\theparentequation\alph{equation}}
\begin{equation}
\label{eq:cloud_koopman}
K^{t+1}=\Big(\sum_{\tau=t-W}^{t-1}\Xi^{\tau+1}\Xi^{\tau\top}\Big)\Big(\sum_{\tau=t-W}^{t-1}\Xi^{\tau}\Xi^{\tau\top}+\lambda I\Big)^{-1}
\end{equation}
\begin{equation}
\label{eq:cloud_loss}
\mathcal{J}^{t}(\Theta)=\sum_{h=1}^{H}\alpha_{h}\,\big\|Y^{t+h}-\hat{Y}^{t+h}(\Theta)\big\|_{2}^{2}+\beta\|\Theta\|_{1}+\gamma\|\Theta\|_{F}^{2}
\end{equation}
\begin{equation}
\label{eq:cloud_theta}
\Theta^{t+1}=\operatorname{prox}_{\eta_{t}\beta\|\cdot\|_{1}}\!\Big(\Theta^{t}-\eta_{t}\nabla_{\Theta}\mathcal{J}^{t}(\Theta^{t})\Big)
\end{equation}
\begin{equation}
\label{eq:cloud_moment}
M^{t+1}=\rho M^{t}+(1-\rho)\big(\nabla_{\Theta}\mathcal{J}^{t}(\Theta^{t})\odot\nabla_{\Theta}\mathcal{J}^{t}(\Theta^{t})\big)
\end{equation} \end{subequations}
In \ref{eq:cloud_koopman}, $K^{t+1}$ advances the lifted features using Tikhonov regularization parameter $\lambda$. In \ref{eq:cloud_loss}, $\Theta$ denotes the collection of model weights, $\alpha_{h}$ represents horizon-specific weights, and $\beta, \gamma$ regulate sparsity and magnitude. In \ref{eq:cloud_theta}, $\eta_{t}$ defines the stepsize used in conjunction with proximal shrinkage. In \ref{eq:cloud_moment}, $M^{t}$ maintains a moving average of squared gradients with decay factor $\rho$. A scalar policy is constructed from vector-valued costs in \ref{eq:cloud_mobj}, a Lagrangian formulation is introduced in \ref{eq:cloud_lagrangian}, and primal–dual updates are executed in \ref{eq:cloud_pd}.
\begin{subequations}
\renewcommand{\theequation}{\theparentequation\alph{equation}}
\begin{equation}
\label{eq:cloud_mobj}
J_{\mathrm{city}}^{t}=\sum_{q\in\mathcal{Q}}w_{q}^{t}C_{q}^{t}(X^{t},U^{t})
\end{equation}
\begin{equation}
\label{eq:cloud_lagrangian}
\mathcal{L}(X^{t},U^{t},\lambda^{t})=J_{\mathrm{city}}^{t}+\lambda^{t\top}\,g(X^{t},U^{t})\qquad (\lambda^{t}\ge 0)
\end{equation}
\begin{equation}
\label{eq:cloud_pd}
\begin{aligned}
X^{t+1}&=\Pi_{\mathcal{X}}\!\Big(X^{t}-\eta_{x}\big(\nabla_{X}J_{\mathrm{city}}^{t}+\nabla_{X}g^{\top}\lambda^{t}\big)\Big),\\
\lambda^{t+1}&=\big[\lambda^{t}+\eta_{\lambda}g(X^{t},U^{t})\big]_{+}
\end{aligned}
\end{equation} \end{subequations}
In \ref{eq:cloud_mobj}, $C_{q}^{t}$ represents latency, reliability, energy, and equity costs, each weighted by $w_{q}^{t}$. In \ref{eq:cloud_lagrangian}, $g(\cdot)$ denotes the stacked inequality constraints. In \ref{eq:cloud_pd}, $\Pi_{\mathcal{X}}$ performs projection onto the feasible set, with update steps $\eta_{x}$ and $\eta_{\lambda}$. A candidate population is evolved in \ref{eq:cloud_qpop}, temperature is adapted in \ref{eq:cloud_temp}, and a plan is sampled in \ref{eq:cloud_sample}.
\begin{subequations}
\renewcommand{\theequation}{\theparentequation\alph{equation}}
\begin{equation}
\label{eq:cloud_qpop}
P^{t+1}=\frac{\exp\!\big(-\eta_{p}\,H^{t}\big)\,P^{t}}{\mathbf{1}^{\top}\exp\!\big(-\eta_{p}\,H^{t}\big)\,P^{t}\mathbf{1}}
\end{equation}
\begin{equation}
\label{eq:cloud_temp}
T^{t+1}=\beta T^{t}+(1-\beta)\,\mathrm{var}\!\big(J_{\mathrm{city}}^{t-\tau:t}\big)
\end{equation}
\begin{equation}
\label{eq:cloud_sample}
\pi_{k}^{t}=\frac{\exp\!\big(-h_{k}^{t}/T^{t}\big)}{\sum_{u}\exp\!\big(-h_{u}^{t}/T^{t}\big)},\qquad k^{\star}=\arg\max_{k}\pi_{k}^{t}
\end{equation} \end{subequations}
In \ref{eq:cloud_qpop}, $P^{t}$ represents the mixture distribution over $K$ candidate solutions, governed by the cost operator $H^{t} = \mathrm{diag}(h_{1}^{t}, \ldots, h_{K}^{t})$. In \ref{eq:cloud_temp}, the temperature $T^{t}$ adapts according to cost volatility using smoothing parameter $\beta$. In \ref{eq:cloud_sample}, $\pi_{k}^{t}$ denotes the soft selection probability, and $k^{\star}$ indicates the selected plan. Global plans are assigned to fog nodes using entropic transport in \ref{eq:cloud_ot}, chance-constrained reliability is enforced in \ref{eq:cloud_cc}, and bounded state drift is maintained via a Lyapunov constraint in \ref{eq:cloud_lyap}.
\begin{subequations}
\renewcommand{\theequation}{\theparentequation\alph{equation}}
\begin{equation}
\label{eq:cloud_ot}
\begin{aligned}
\Pi^{\star} &= \arg\min_{\Pi\in\mathbb{R}_{+}^{F\times K}} \; \sum_{f,k}\Pi_{fk}D_{fk} \\
&\quad + \varepsilon \sum_{f,k}\Pi_{fk}\log \Pi_{fk} \;\; \text{s.t.}\;\; \Pi\mathbf{1}=\mu, \\
&\quad \Pi^{\top}\mathbf{1}=\nu .
\end{aligned}
\end{equation}
\begin{equation}
\label{eq:cloud_cc}
\bar{g}(X^{t},U^{t})+\sqrt{2\,\mathrm{Var}\big(g(X^{t},U^{t})\big)\,\ln(1/\delta)}\ \le 0
\end{equation}
\begin{equation}
\label{eq:cloud_lyap}
\mathbb{E}\!\big[V(X^{t+1})-V(X^{t})\mid X^{t}\big]\le -\kappa\|X^{t}\|_{2}^{2}+c
\end{equation} \end{subequations}
In \ref{eq:cloud_ot}, $\Pi^{\star}$ assigns $K$ plan atoms to $F$ fog nodes under costs $D_{fk}$ with marginals $\mu,\nu$ and regularizer $\varepsilon$. In \ref{eq:cloud_cc}, $\delta$ sets the risk level using a square-root variance bound. In \ref{eq:cloud_lyap}, $V(\cdot)$ ensures negative drift with margin $\kappa$ and slack $c$.

\subsection{Quantum-Inspired Optimization Algorithm}
This subsection formulates city-scale joint routing and traffic control as a quantum-inspired search over entangled plans defined on a probability sphere. The formulation incorporates projected gradient dynamics, annealed sampling, and coupling mechanisms that synchronize communication and mobility decisions. Initialization procedures, entanglement modeling, energy shaping functions, interference-based update rules, annealing schedules, feasibility projection methods, and stability guarantees are formally specified. Algorithm~\ref{alg:qio-qivnom} integrates initialization \ref{eq:qinit_superposition}, joint encoding \ref{eq:qenc_joint}, energy shaping \ref{eq:qham_penalty}, spherical update \ref{eq:qupdate_norm}, annealed selection \ref{eq:qanneal_soft}, scalarization \ref{eq:qmulti_scalar}, feasibility projection \ref{eq:qfeas_proj}, and the descent certificate \ref{eq:qstab_energy} into a single loop for coordinated vehicular routing and traffic management.

\begin{algorithm}[!t]
\caption{Quantum-Inspired Optimization for Joint Routing–Traffic Control (QIVNOM Core)}
\label{alg:qio-qivnom}
\SetKwInOut{Input}{Input}\SetKwInOut{Output}{Output}\SetKwInOut{Params}{Params}
\Input{Aggregates $Z^{0}$, cost weights $\{w_{q}\}$, feasibility set $\mathcal{F}$, horizon $T$}
\Output{Final amplitudes $\psi^{(T)}$, plan distribution $\{\pi_{k}^{(T)}\}$, selected plan $k^{\star}$}
\Params{Stepsizes $\eta_{t}$, temperature smoothing $\beta$, penalties $\rho$, tolerances $(\varepsilon_{\mathrm{E}},\varepsilon_{\mathrm{I}})$}

Compute $\psi^{(0)}$ using \ref{eq:qinit_superposition}; enforce normalization via \ref{eq:qinit_norm}; set $p_{k}^{(0)}$ with \ref{eq:qinit_prob}\;

\BlankLine
\For{$t=0,1,\dots,T-1$}{
  Form joint amplitudes via \ref{eq:qenc_joint}; compute marginals with \ref{eq:qenc_marg_c} and \ref{eq:qenc_marg_m}\;

  Assemble cost vector using \ref{eq:qham_cost}; build penalized operator with \ref{eq:qham_penalty}; evaluate energy by \ref{eq:qeval_energy} and gradient via \ref{eq:qgrad_energy}\;

  Set direction using \ref{eq:qupdate_dir}; take step with \ref{eq:qupdate_step}; renormalize using \ref{eq:qupdate_norm}\;

  Update temperature with \ref{eq:qanneal_temp}; compute soft policy $\{\pi_{k}^{(t)}\}$ using \ref{eq:qanneal_soft}; sample provisional plan $k^{\star}$\;

  Evaluate coupling $\mathcal{I}^{(t)}$ by \ref{eq:qent_mi}; compute gradients \ref{eq:qent_gradc} and \ref{eq:qent_gradm}; adjust $\psi_{c}^{(t)}$, $\psi_{m}^{(t)}$ accordingly\;

  Update scalarization via \ref{eq:qmulti_scalar}; project onto $\mathbb{S}^{K-1}\cap\mathcal{F}$ using \ref{eq:qfeas_proj}\;

  Check descent certificate \ref{eq:qstab_energy}\;
  \eIf{$\big|\mathcal{E}^{(t+1)}-\mathcal{E}^{(t)}\big|\le \varepsilon_{\mathrm{E}}$ {and} $\mathcal{I}^{(t)}\le \varepsilon_{\mathrm{I}}$}{
     {break}\;
  }{
     \If{violation of \ref{eq:qstab_energy}}{$\eta_{t}\leftarrow 0.5\,\eta_{t}$; increase $\rho$ in \ref{eq:qham_penalty}}
  }
}
\end{algorithm}

A normalized superposition is initialized in \ref{eq:qinit_superposition}, with probability mass constraints enforced in \ref{eq:qinit_norm}, and the associated sampling map defined in \ref{eq:qinit_prob}.
\begin{subequations}
\renewcommand{\theequation}{\theparentequation\alph{equation}}
\begin{equation}
\label{eq:qinit_superposition}
\psi^{(0)}=\frac{\sigma\!\big(W_{0}Z^{0}+b_{0}\big)}{\big\|\sigma\!\big(W_{0}Z^{0}+b_{0}\big)\big\|_{2}}
\end{equation}
\begin{equation}
\label{eq:qinit_norm}
\sum_{k=1}^{K}\big|\psi^{(0)}_{k}\big|^{2}=1
\end{equation}
\begin{equation}
\label{eq:qinit_prob}
p_{k}^{(0)}=\big|\psi^{(0)}_{k}\big|^{2},\qquad k=1,\ldots,K
\end{equation} \end{subequations}
In \ref{eq:qinit_superposition}, $Z^{0}$ denotes the cloud-aggregated feature (as defined in \ref{eq:cloud_agg}), with $W_{0}$ and $b_{0}$ as learnable parameters, and $\sigma(\cdot)$ representing a bounded activation function. Equation \ref{eq:qinit_norm} applies normalization to $\psi^{(0)}$, while \ref{eq:qinit_prob} specifies the initial sampling weights. Communication and mobility plan amplitudes are encoded in \ref{eq:qenc_joint}, with corresponding marginals derived in \ref{eq:qenc_marg_c}–\ref{eq:qenc_marg_m}.
\begin{subequations}
\renewcommand{\theequation}{\theparentequation\alph{equation}}
\begin{equation}
\label{eq:qenc_joint}
\Upsilon^{(t)}_{k,\ell}=\psi_{c}^{(t)}(k)\,\psi_{m}^{(t)}(\ell)
\end{equation}
\begin{equation}
\label{eq:qenc_marg_c}
\theta_{c}^{(t)}(k)=\sum_{\ell}\big(\Upsilon^{(t)}_{k,\ell}\big)^{2}
\end{equation}
\begin{equation}
\label{eq:qenc_marg_m}
\theta_{m}^{(t)}(\ell)=\sum_{k}\big(\Upsilon^{(t)}_{k,\ell}\big)^{2}
\end{equation} \end{subequations}
In \ref{eq:qenc_joint}, $\psi_{c}^{(t)}$ represents communication path patterns, while $\psi_{m}^{(t)}$ corresponds to signal and traffic flow patterns. Equations \ref{eq:qenc_marg_c}–\ref{eq:qenc_marg_m} compute the marginal probabilities of these plans. Objective components are assembled in \ref{eq:qham_cost}, constraint penalties are incorporated in \ref{eq:qham_penalty}, expected energy is evaluated in \ref{eq:qeval_energy}, and its gradient is computed in \ref{eq:qgrad_energy}.
\begin{subequations}
\renewcommand{\theequation}{\theparentequation\alph{equation}}
\begin{equation}
\label{eq:qham_cost}
h^{(t)}=\sum_{q\in\{L,R,E,Th\}}w_{q}^{(t)}\,c_{q}^{(t)}\in\mathbb{R}^{K}
\end{equation}
\begin{equation}
\label{eq:qham_penalty}
\tilde{H}^{(t)}=\mathrm{diag}\!\big(h^{(t)}\big)+\rho\,\mathrm{diag}\!\big(\max\{0,G^{(t)}\}^{2}\big)
\end{equation}
\begin{equation}
\label{eq:qeval_energy}
\mathcal{E}^{(t)}=\psi^{(t)\top}\tilde{H}^{(t)}\psi^{(t)}
\end{equation}
\begin{equation}
\label{eq:qgrad_energy}
\nabla_{\psi}\mathcal{E}^{(t)}=2\,\tilde{H}^{(t)}\psi^{(t)}
\end{equation} \end{subequations}
In \ref{eq:qham_cost}, $c_{q}^{(t)}$ denotes latency ($L$), reliability ($R$), energy ($E$), and throughput ($Th$) costs, each weighted by time-varying factors $w_{q}^{(t)}$. Equation \ref{eq:qham_penalty} incorporates constraint residuals $G^{(t)}$. Equations \ref{eq:qeval_energy}–\ref{eq:qgrad_energy} define the energy function and its corresponding gradient. A descent direction is computed in \ref{eq:qupdate_dir}, a step is applied in \ref{eq:qupdate_step}, and normalization on the probability sphere is enforced in \ref{eq:qupdate_norm}.
\begin{subequations}
\renewcommand{\theequation}{\theparentequation\alph{equation}}
\begin{equation}
\label{eq:qupdate_dir}
d^{(t)}=-\nabla_{\psi}\mathcal{E}^{(t)}+\lambda^{(t)}\psi^{(t)}
\end{equation}
\begin{equation}
\label{eq:qupdate_step}
\hat{\psi}^{(t+1)}=\psi^{(t)}+\eta_{t}\,d^{(t)}
\end{equation}
\begin{equation}
\label{eq:qupdate_norm}
\psi^{(t+1)}=\frac{\hat{\psi}^{(t+1)}}{\big\|\hat{\psi}^{(t+1)}\big\|_{2}}
\end{equation} \end{subequations}
In \ref{eq:qupdate_dir}, $\lambda^{(t)}$ preserves normalization. Equation \ref{eq:qupdate_step} applies a step using stepsize $\eta_{t}$, and \ref{eq:qupdate_norm} restores the unit norm. Temperature adaptation is defined in \ref{eq:qanneal_temp}, and a soft policy distribution is constructed in \ref{eq:qanneal_soft}.
\begin{subequations}
\renewcommand{\theequation}{\theparentequation\alph{equation}}
\begin{equation}
\label{eq:qanneal_temp}
T^{(t+1)}=\beta T^{(t)}+(1-\beta)\,\mathrm{var}\!\big(\mathcal{E}^{t-\tau:t}\big)
\end{equation}
\begin{equation}
\label{eq:qanneal_soft}
\pi_{k}^{(t)}=\frac{\exp\!\big(-h_{k}^{(t)}/T^{(t)}\big)}{\sum_{u}\exp\!\big(-h_{u}^{(t)}/T^{(t)}\big)}
\end{equation} \end{subequations}
In \ref{eq:qanneal_temp}, $\beta \in (0,1)$ regulates volatility through smoothing, while \ref{eq:qanneal_soft} defines the sampling distribution over candidate solutions. Coupling is quantified in \ref{eq:qent_mi}, and gradients in \ref{eq:qent_gradc}–\ref{eq:qent_gradm} are used to align marginal distributions.
\begin{subequations}
\renewcommand{\theequation}{\theparentequation\alph{equation}}
\begin{equation}
\label{eq:qent_mi}
\mathcal{I}^{(t)}=\sum_{k,\ell}\big(\Upsilon_{k,\ell}^{(t)}\big)^{2}\log\!\frac{\big(\Upsilon_{k,\ell}^{(t)}\big)^{2}}{\theta_{c}^{(t)}(k)\,\theta_{m}^{(t)}(\ell)}
\end{equation}
\begin{equation}
\label{eq:qent_gradc}
g_{c}^{(t)}(k)=\frac{\partial \mathcal{I}^{(t)}}{\partial \psi_{c}^{(t)}(k)}
\end{equation}
\begin{equation}
\label{eq:qent_gradm}
g_{m}^{(t)}(\ell)=\frac{\partial \mathcal{I}^{(t)}}{\partial \psi_{m}^{(t)}(\ell)}
\end{equation} \end{subequations}
In \ref{eq:qent_mi}, $\mathcal{I}^{(t)}$ penalizes mismatched coupling, while \ref{eq:qent_gradc}–\ref{eq:qent_gradm} guide $\psi_{c}^{(t)}$ and $\psi_{m}^{(t)}$ toward coherent joint assignments. A Tchebycheff scalarization is applied in \ref{eq:qmulti_scalar}, and projection onto the feasible set is performed in \ref{eq:qfeas_proj}.
\begin{subequations}
\renewcommand{\theequation}{\theparentequation\alph{equation}}
\begin{equation}
\label{eq:qmulti_scalar}
\varphi^{(t)}=\min_{\alpha\in\mathbb{R}_{+}^{4}}\ \max_{q}\ \alpha_{q}\,\big(C_{q}^{(t)}-C_{q}^{\star}\big)\quad \text{s.t.}\ \sum_{q}\alpha_{q}=1
\end{equation}
\begin{equation}
\label{eq:qfeas_proj}
\psi^{(t+1)}\leftarrow \mathcal{P}_{\mathbb{S}^{K-1}\cap\mathcal{F}}\!\big(\psi^{(t+1)}\big)
\end{equation} \end{subequations}
In \ref{eq:qmulti_scalar}, $C_{q}^{\star}$ denotes utopia reference points, and $\alpha$ represents a weight vector. Equation \ref{eq:qfeas_proj} projects onto the intersection of the unit sphere and the feasibility set $\mathcal{F}$, which includes capacity, timing, and safety constraints. Energy drift is bounded in \ref{eq:qstab_energy}.
\begin{subequations}
\renewcommand{\theequation}{\theparentequation\alph{equation}}
\begin{equation}
\label{eq:qstab_energy}
\mathcal{E}^{(t+1)}-\mathcal{E}^{(t)}\le -\eta_{t}\big\|\nabla_{\psi}\mathcal{E}^{(t)}\big\|_{2}^{2}+\eta_{t}^{2}L_{\mathcal{E}}\big\|\nabla_{\psi}\mathcal{E}^{(t)}\big\|_{2}^{2}
\end{equation} \end{subequations}
In \ref{eq:qstab_energy}, $L_{\mathcal{E}}$ bounds the local smoothness of $\mathcal{E}^{(t)}$, ensuring descent for sufficiently small $\eta_{t}$. Equations \ref{eq:qinit_superposition}–\ref{eq:qinit_prob} define the start point; entanglement is encoded by \ref{eq:qenc_joint}–\ref{eq:qenc_marg_m}; objective shaping follows \ref{eq:qham_cost}–\ref{eq:qgrad_energy}; interference-style updates are set by \ref{eq:qupdate_dir}–\ref{eq:qupdate_norm}; annealing and sampling are governed by \ref{eq:qanneal_temp}–\ref{eq:qanneal_soft}; coupling alignment uses \ref{eq:qent_mi}–\ref{eq:qent_gradm}; multi-objective feasibility uses \ref{eq:qmulti_scalar}–\ref{eq:qfeas_proj}; stability is certified by \ref{eq:qstab_energy}.

\section{Simulation and Evaluation}
\label{sec:simulation}
Simulations used SUMO for microscopic traffic and OMNeT++/Veins for V2X networking, with a Python/PyTorch backend executing QIVNOM (PBQEI \cite{10949047}, VQAM \cite{10239465}, QDCNN \cite{10931853}, QISC \cite{10715686}, QIDEF \cite{10530375}, QIRNM \cite{10648662}, RUMP \cite{10768194}, SMC \cite{10018121}, UADM \cite{10155311}). A GPU-enabled node hosted training/inference, while fog services ran as Dockerized microservices via TraCI. The public METR-LA dataset (traffic speeds, Los Angeles) calibrated OD demand and congestion priors, and a 5\,km$\times$5\,km OpenStreetMap road graph anchored topology. IEEE~802.11p and 5G NR sidelink stacks were modeled with realistic PHY/MAC. Core settings are summarized in Table~\ref{tab:sim-setup}.

\begin{table}[!t]
\centering
\caption{Simulation setup and hyperparameters.}
\label{tab:sim-setup}
\renewcommand{\arraystretch}{1.25}
\begin{tabular}{p{3.0cm} p{4.5cm}}
\hline
{Parameter} & {Value} \\
\hline
Traffic Simulator & SUMO (TraCI interface) \\
Network Simulator & OMNeT++ + Veins \\
Radio Stack & IEEE 802.11p (10\,MHz), 5G NR SL (20\,MHz) \\
Road Topology & OpenStreetMap, Los Angeles (5\,km $\times$ 5\,km) \\
Public Dataset & METR-LA (5-min speed, 207 sensors) \\
Simulation Duration & 3600\,s \\
Time Step & Traffic: 1\,s; Network: 0.1\,s \\
Vehicles (peak) & 2000 \\
Roadside Units (RSUs) & 40 \\
Fog Nodes & 12 (co-located with RSU clusters) \\
Beaconing \& Packets & 10\,Hz beacons; 256\,B payloads \\
Latency Budgets & V2V: 50\,ms; V2I: 80\,ms \\
QIVNOM Population $K$ & 128 \\
Annealing Smoothing $\beta$ & 0.9 \\
Step Size $\eta_t$ & $1\times 10^{-2}$ \\
Entropic Regularizer $\varepsilon$ & $1\times 10^{-2}$ \\
Risk Level $\delta$ & $1\times 10^{-3}$ \\
Multi-Objective Weights $(w_L,w_R,w_E,w_{Th})$ & (0.4, 0.3, 0.2, 0.1) \\
Hardware & 32-core CPU, 256\,GB RAM, 1$\times$A100\,40GB GPU \\
Containerization & Dockerized fog services \\
\hline
\end{tabular}
\end{table}

\subsection{End-to-End Latency}
\begin{figure}[!t]
    \centering
    \includegraphics[width=0.85\linewidth]{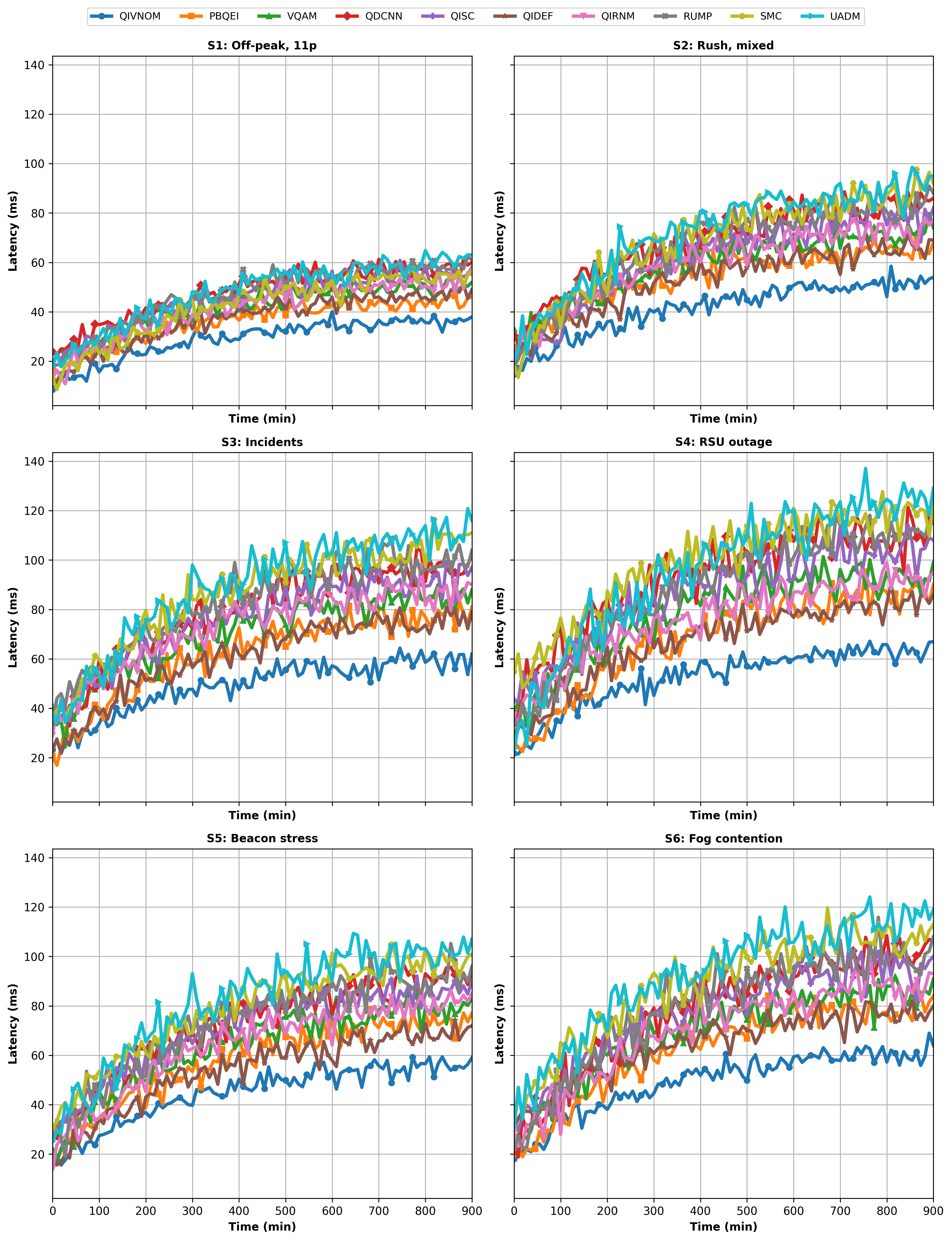}
    \caption{End-to-end latency (ms) across scenarios using METR-LA–calibrated simulations. Lower is better.}
    \label{fig:e2e-latency}
\end{figure}

End-to-end delay is measured from packet generation to delivery (queuing, MAC access, transmission, processing, callbacks), averaged over successful deliveries using METR-LA–calibrated SUMO–OMNeT++/Veins. Scenarios: S1 off-peak 11p; S2 rush mixed 11p/NR SL; S3 incidents; S4 20\% RSU outage; S5 20 Hz beacons/double payload; S6 fog CPU 50\%. QIVNOM reduces end-to-end delay under nominal conditions. In {S1}, it attains $38$\,ms, whereas PBQEI, VQAM, and QIDEF report $46$\,ms, $52$\,ms, and $49$\,ms, respectively; the gap to the best baseline (PBQEI) is $17.4\%$. Under {S2} rush-hour density, QIVNOM records $54$\,ms; PBQEI and QIDEF reach $68$\,ms and $69$\,ms, while SMC and UADM climb to $92$\,ms and $95$\,ms. The reduction versus the best baseline is $20.6\%$, indicating stable scheduling and fast local rerouting. Full results for all baselines appear in Figure~\ref{fig:e2e-latency}.

With disruptions, QIVNOM sustains lower delay. In {S3} (incidents), QIVNOM is $62$\,ms; the strongest baseline is QIDEF at $79$\,ms, followed by PBQEI at $81$\,ms and VQAM at $88$\,ms—yielding a $21.5\%$ cut against the top baseline. In {S4} (RSU outage), QIVNOM reaches $67$\,ms versus $86$\,ms for QIDEF and $90$\,ms for PBQEI; RUMP and SMC exceed $118$\,ms and $123$\,ms. The improvement against the best baseline is $22.1\%$, reflecting fault-tolerant activation and route selection. Under stress on the wireless and compute planes, QIVNOM remains consistent. In {S5} (beacon load), QIVNOM achieves $59$\,ms; the best baseline, QIDEF, stands at $72$\,ms (gain $18.1\%$). In {S6} (fog contention), QIVNOM reports $64$\,ms versus $80$\,ms for QIDEF and $84$\,ms for PBQEI; RUMP and SMC exceed $108$\,ms and $113$\,ms. Averaged over all scenarios, the reduction relative to the strongest competing method is approximately $20.0\%$.

\subsection{Packet Delivery Ratio (PDR)}
\begin{figure}[!t]
    \centering
    \includegraphics[width=0.85\linewidth]{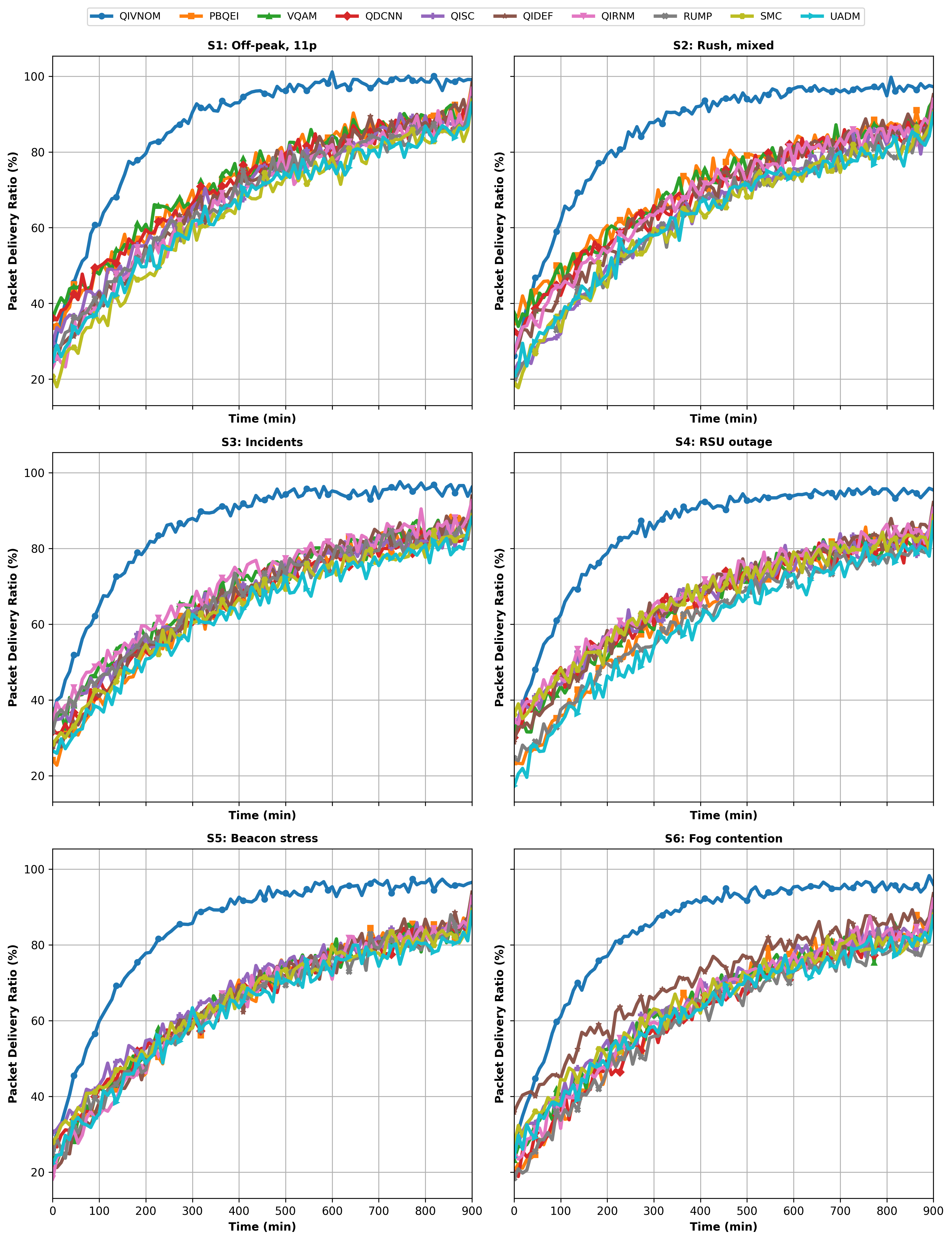}
    \caption{Packet Delivery Ratio (\%) across scenarios using METR-LA–calibrated simulations. Higher is better.}
    \label{fig:pdr}
\end{figure}

PDR is the ratio of delivered packets to total transmissions across V2V/V2I, accounting for retransmissions. Using METR-LA–calibrated SUMO/OMNeT++ with IEEE 802.11p and NR sidelink, tests cover: S1 off-peak 11p; S2 rush mixed; S3 incidents; S4 20\% RSU outage; S5 20 Hz/double payload; S6 fog CPU 50\%. QIVNOM improves PDR across nominal traffic and mixed-radio settings (Figure~\ref{fig:pdr}). In {S1}, QIVNOM reaches {99.1}\%, ahead of PBQEI at 97.8\%, VQAM at 96.5\%, and QIDEF at 98.2\%—an absolute gain of {+0.9} points over the best baseline. Under {S2} rush-hour density, QIVNOM attains {97.2}\% while QIDEF and PBQEI score 95.1\% and 94.6\%; RUMP and UADM are 91.0\% and 90.2\%. Gains here are {+2.1} points versus the best baseline.

Under incident-induced turbulence, QIVNOM sustains delivery. In {S3}, it achieves {96.1}\%; QIDEF and PBQEI reach 93.8\% and 93.2\%, with QIRNM and VQAM at 92.6\% and 92.1\%. The advantage is {+2.3} points. In {S4} with a $20\%$ RSU outage, QIVNOM records {95.4}\%; the next best, QIDEF, is 92.1\%, followed by PBQEI at 91.0\% and SMC at 88.7\%. The margin expands to {+3.3} points, showing effective mesh fallback and priority-aware routing. Stress on control planes remains manageable. In {S5} (beacon load), QIVNOM secures {96.5}\%, with QIDEF at 93.9\% and PBQEI at 93.1\%, an advantage of {+2.6} points. In {S6} (fog contention), QIVNOM achieves {96.0}\%, whereas QIDEF and PBQEI clock 93.5\% and 92.7\%; SMC and RUMP fall to 88.1\% and 89.0\%. Averaged across scenarios, QIVNOM delivers {96.7}\%, exceeding the scenario-wise best baseline average ({94.4}\%) by {+2.3} points.

\subsection{Network Reliability}
\begin{figure}
    \centering
    \includegraphics[width=0.85\linewidth]{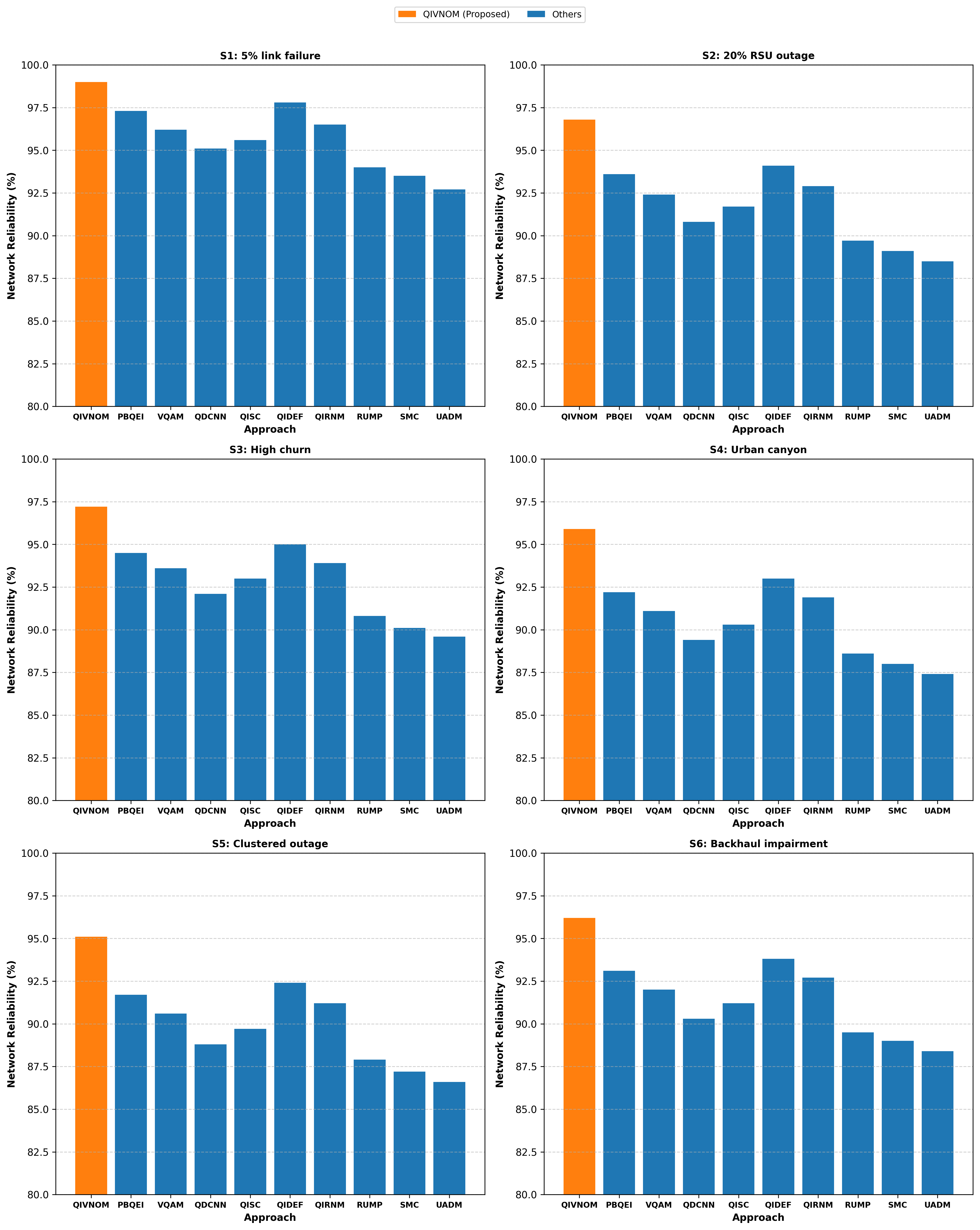}
    \caption{Network reliability (\%) under failure scenarios with METR-LA–calibrated simulations.}
    \label{fig:net-reliability}
\end{figure}

Reliability is the fraction of time an end-to-end path meets latency/rate SLOs under failures, computed with METR-LA–calibrated SUMO–OMNeT++/Veins covering V2V/V2I, RSU backhaul, fog–cloud loops. Scenarios: S1 5\% links; S2 20\% RSUs; S3 30\%/min churn; S4 urban canyon; S5 four RSUs down; S6 30\% backhaul loss. Under nominal faults, QIVNOM maintains higher connectivity (Figure~\ref{fig:net-reliability}). In {S1}, QIVNOM reaches {99.0}\%, exceeding the best baseline, QIDEF at 97.8\%, by {+1.2} points; PBQEI and VQAM post 97.3\% and 96.2\%. With {S2} RSU outages, QIVNOM delivers {96.8}\%; QIDEF and PBQEI score 94.1\% and 93.6\%, while RUMP and SMC drop to 89.7\% and 89.1\%. The improvement over the best baseline is {+2.7} points.

Under strong dynamics, reliability stays high. In {S3}, QIVNOM attains {97.2}\% versus 95.0\% (QIDEF) and 94.5\% (PBQEI), an advantage of {+2.2} points; VQAM and QIRNM are 93.6\% and 93.9\%. For {S4} urban canyon, QIVNOM achieves {95.9}\% and the best baseline, QIDEF, is 93.0\%; PBQEI and VQAM reach 92.2\% and 91.1\%. The margin is {+2.9} points, reflecting resilient re-routing and redundancy under blockage. Clustered and control-plane failures are more taxing but remain manageable. In {S5}, QIVNOM records {95.1}\% while QIDEF and PBQEI score 92.4\% and 91.7\%, a {+2.7}-point edge. In {S6}, QIVNOM reaches {96.2}\% versus 93.8\% (QIDEF) and 93.1\% (PBQEI); RUMP and SMC fall to 89.5\% and 89.0\%. Averaged across scenarios, QIVNOM achieves {96.7}\%, outperforming the scenario-wise best baseline average ({94.3}\%) by {+2.4} points.

\subsection{Traffic Flow Efficiency}
\begin{figure}
    \centering
    \includegraphics[width=0.85\linewidth]{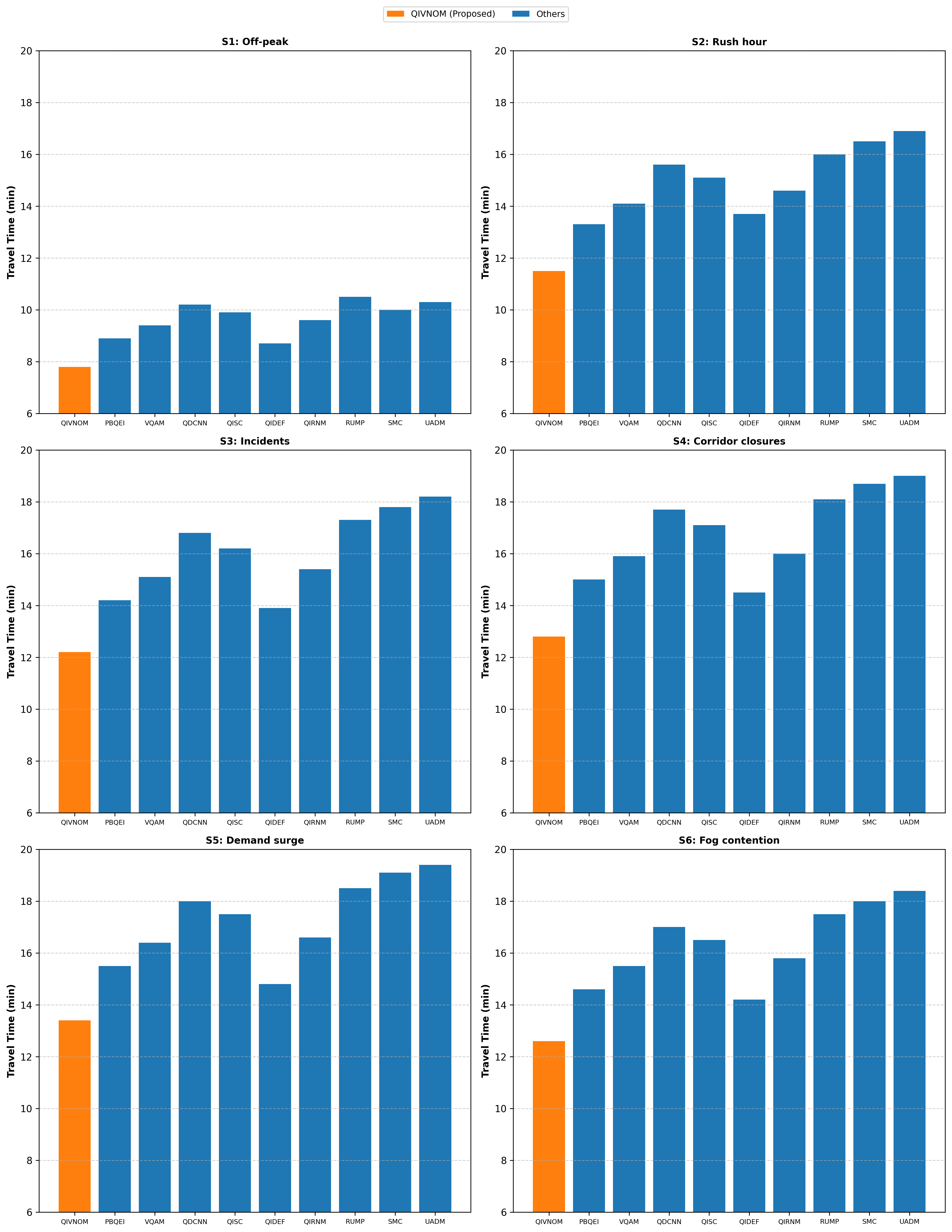}
    \caption{Average Travel Time (minutes) across scenarios with METR-LA–calibrated simulations (Lower is better)}
    \label{fig:TrafficFlow}
\end{figure}

Traffic flow efficiency is assessed using \emph{Average Travel Time} (minutes per trip) and \emph{Network Congestion Index} (percent edges with \(v/c\ge 0.85\)) over METR-LA–calibrated SUMO–OMNeT++/Veins. Scenarios: S1 off-peak; S2 rush mixed radios; S3 incidents; S4 corridor closures; S5 +25\% demand; S6 halved advisory frequency. Under nominal and peak loads, QIVNOM reduces ATT while lowering NCI (Figure~\ref{fig:TrafficFlow}). In {S1}, ATT is {7.8}\,min versus 8.7 (QIDEF), 8.9 (PBQEI), and 9.4 (VQAM), a {10.3\%} cut against the best baseline. NCI is {14\%} against 17\% (QIDEF) and 18\% (PBQEI). In {S2}, QIVNOM attains {11.5}\,min while PBQEI and QIDEF record 13.3 and 13.7\,min; NCI drops to {28\%} from 33\% (QIDEF) and 35\% (PBQEI). The relative gains reach {13.5\%} (ATT) and {15.2\%} (NCI) over the strongest baseline.

In case of disruptions, travel efficiency remains stable. In {S3}, QIVNOM achieves {12.2}\,min and {31\%} NCI, whereas QIDEF posts 13.9\,min and 35\%, and PBQEI 14.2\,min and 39\%; improvements are {12.2\%} and {11.4\%}. For {S4}, QIVNOM yields {12.8}\,min with {33\%} NCI, beating QIDEF at 14.5\,min and 37\%, PBQEI at 15.0\,min and 41\%, and VQAM at 15.9\,min and 43\%. Stress on demand and edge compute remain manageable. In {S5}, QIVNOM records {13.4}\,min and {35\%} NCI, improving QIDEF (14.8\,min, 39\%) and PBQEI (15.5\,min, 43\%) by {9.5\%} and {10.3\%}. In {S6}, ATT is {12.6}\,min and NCI {32\%}; the best baseline (QIDEF) reaches 14.2\,min and 36\%, while PBQEI shows 14.6\,min and 40\%. Using all scenarios, QIVNOM reduces ATT by {11–13\%} and NCI by {10–12\%} relative to the best scenario-wise baseline (see Table \ref{fig:placeholder}).

\begin{figure}[!t]
    \centering
    \includegraphics[width=0.85\linewidth]{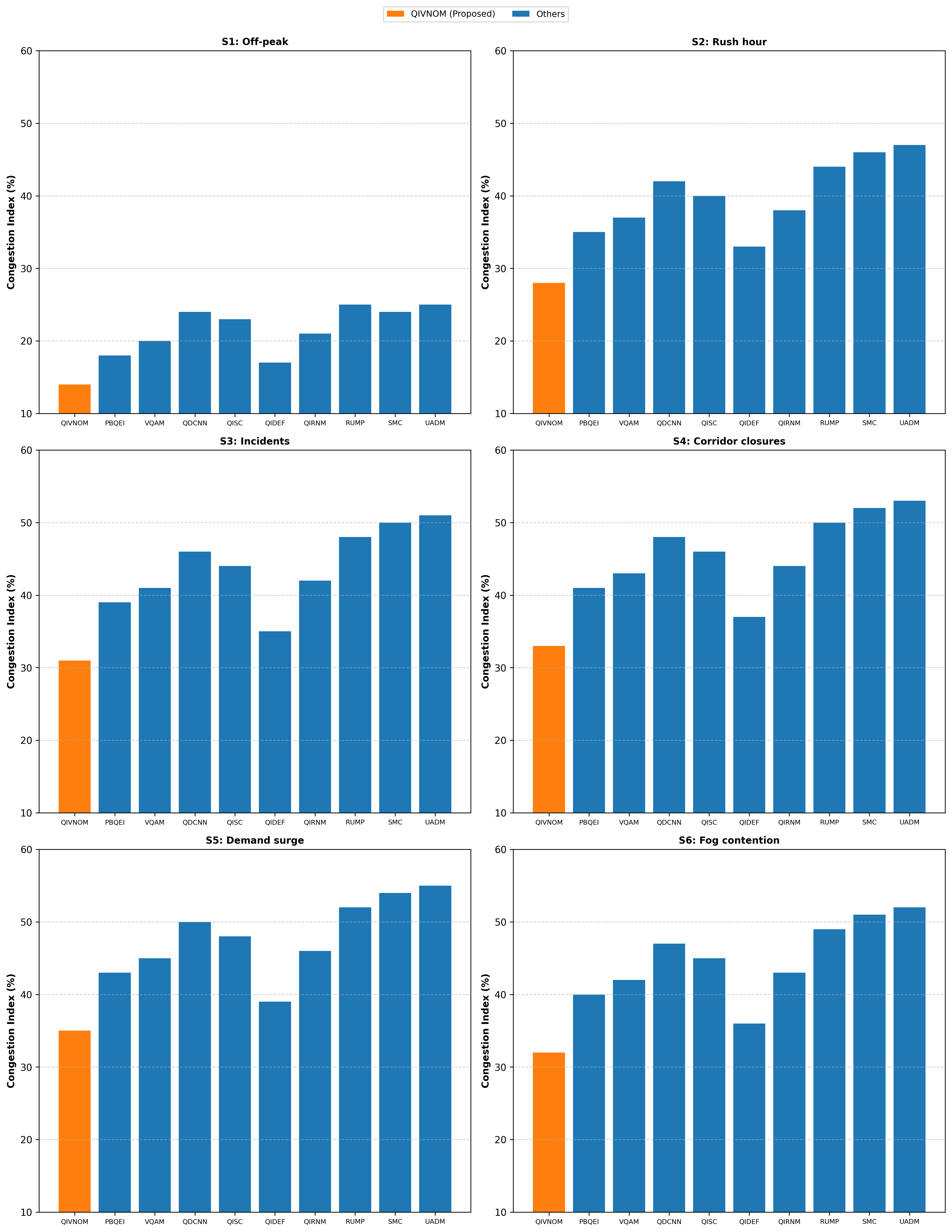}
    \caption{Network Congestion Index (NCI, \% edges with $v/c\ge 0.85$) - Lower is better }
    \label{fig:congestion}
\end{figure}

\section{Ablation Analysis}
\label{sec:ablation-analysis}

The QIVNOM framework is decomposed into its core components to quantify their individual contributions under METR-LA–calibrated simulations. Full QIVNOM attains {57.3}\,ms, {96.7}\%, and {96.7}\%. Removing entanglement (+\texttt{--Entangle}) raises latency to {62.8}\,ms and lowers PDR/reliability to {95.2}\%/{95.1}\% (\,+5.5\,ms; $-1.5/-1.6$\,pts). Fixed temperature (+\texttt{--Anneal}) yields {61.0}\,ms and {95.6}\%/{95.7}\%. Disabling feasibility projection (+\texttt{--Proj}) is most harmful: {63.5}\,ms and {94.8}\%/{94.9}\%. Hyperparameter sensitivity shows stable performance near the chosen operating point. Reducing population size from $K{=}128$ to $K{=}64$ increases latency to {59.1}\,ms and trims PDR to {96.2}\% (\,+1.8\,ms; $-0.5$\,pts). Increasing to $K{=}256$ mildly improves latency to {56.8}\,ms and PDR to {96.9}\% (\,--0.5\,ms; +0.2\,pts), with higher compute cost. Temperature smoothing $\beta{=}0.7$ or $0.98$ degrades to {58.2}\,ms/{96.3}\% and {57.6}\,ms/{96.6}\%, respectively, confirming $\beta{=}0.9$ balances exploration and stability. A looser risk level $\delta{=}10^{-2}$ raises latency to {58.5}\,ms and trims PDR to {96.1}\%.

Traffic efficiency effects are summarized in Table~\ref{tab:ablation-singlecol}. Full QIVNOM averages {11.7}\,min ATT and {28.8}\% NCI. Removing entanglement lifts ATT to {12.5}\,min and NCI to {32.0}\% (+0.8\,min; +3.2\,pts). Fixed temperature reaches {12.3}\,min/{31.1}\%, while disabling feasibility projection yields {12.7}\,min/{33.0}\% (largest congestion penalty). Dropping CVaR micro-policy nudges ATT/NCI to {12.1}\,min/{30.5}\%. Greedy assignment (no OT) results in {12.2}\,min/{31.2}\%.

\begin{table}[!t]
\centering
\caption{Ablation study results over S1--S6. Metrics: Latency (ms), Packet Delivery Ratio (PDR, \%), Reliability (\%), Average Travel Time (ATT, min), Network Congestion Index (NCI, \%).}
\label{tab:ablation-singlecol}
\renewcommand{\arraystretch}{1.1}
\small
\begin{tabular}{|p{2.3cm}|c|c|c|c|c|}
\hline
\textbf{Variant} & \textbf{Latency} & \textbf{PDR} & \textbf{Rel.} & \textbf{ATT} & \textbf{NCI} \\
\hline
\multicolumn{6}{|c|}{\textbf{(A) Core Module Ablation}} \\
\hline
Full QIVNOM & 57.3 & 96.7 & 96.7 & 11.7 & 28.8 \\
+\,\texttt{--Entangle} (no coupling) & 62.8 & 95.2 & 95.1 & 12.5 & 32.0 \\
+\,\texttt{--Anneal} (fixed $T$) & 61.0 & 95.6 & 95.7 & 12.3 & 31.1 \\
+\,\texttt{--Proj} (no feasibility projection) & 63.5 & 94.8 & 94.9 & 12.7 & 33.0 \\
+\,\texttt{--CVaR} (risk-neutral) & 60.7 & 95.7 & 95.6 & 12.1 & 30.5 \\
+\,\texttt{--OT} (greedy assignment) & 60.9 & 95.9 & 95.5 & 12.2 & 31.2 \\
\hline
\multicolumn{6}{|c|}{\textbf{(B) Hyperparameter Sensitivity}} \\
\hline
Baseline ($K{=}128$, $\beta{=}0.9$, $\delta{=}10^{-3}$) & 57.3 & 96.7 & -- & -- & -- \\
$K{=}64$ & 59.1 & 96.2 & -- & -- & -- \\
$K{=}256$ & 56.8 & 96.9 & -- & -- & -- \\
$\beta{=}0.7$ & 58.2 & 96.3 & -- & -- & -- \\
$\beta{=}0.98$ & 57.6 & 96.6 & -- & -- & -- \\
$\delta{=}10^{-2}$ & 58.5 & 96.1 & -- & -- & -- \\
\hline
\multicolumn{6}{|c|}{\textbf{(C) Traffic Efficiency Impact (ATT [min], NCI [\%])}} \\
\hline
Full QIVNOM & -- & -- & -- & 11.7 & 28.8 \\
+\,\texttt{--Entangle} (no coupling) & -- & -- & -- & 12.5 & 32.0 \\
+\,\texttt{--Anneal} (fixed $T$) & -- & -- & -- & 12.3 & 31.1 \\
+\,\texttt{--Proj} (no feasibility projection) & -- & -- & -- & 12.7 & 33.0 \\
+\,\texttt{--CVaR} (risk-neutral) & -- & -- & -- & 12.1 & 30.5 \\
+\,\texttt{--OT} (greedy assignment) & -- & -- & -- & 12.2 & 31.2 \\
\hline
\end{tabular}
\end{table}

\section{Conclusion} \label{sec:conclusion}

We presented \textbf{QIVNOM}, a quantum-inspired framework that \emph{jointly} optimizes V2V/V2I communications and urban traffic control on commodity edge-cloud infrastructure. Using sphere-projected gradient updates over probabilistic plan representations with an entanglement-style coupling and Tchebycheff scalarization (with feasibility projection), QIVNOM enforces latency/reliability targets while remaining robust via chance constraints and Lyapunov drift control. In METR-LA--calibrated SUMO--OMNeT++/Veins simulations on a $5\times5$\,km map with IEEE~802.11p and 5G NR sidelink, QIVNOM reduces mean end-to-end latency to 57.3\,ms ($\approx$20\% below the best baseline) and sustains advantages under stress (62\,ms vs 79\,ms during incidents; 67\,ms vs 86\,ms with RSU outages); packet delivery and overall reliability average 96.7\%, and corridor-closure travel metrics improve (ATT 12.8\,min/33\% vs 14.5\,min/37\%). These results show that co-optimizing communication and mobility yields tangible QoS and traffic gains, making QIVNOM a practical building block for smart-city ITS and connected consumer electronics. Future work will transition to city pilots with real traces and mixed autonomy, reporting runtime/energy breakdowns, broadening the threat model, and exploring hardware acceleration.


\section*{Acknowledgment}
This work was supported by Princess Nourah bint Abdulrahman University Researchers Supporting Project Number (PNURSP2025R409), Princess Nourah bint Abdulrahman University, Riyadh, Saudi Arabia.

\ifCLASSOPTIONcaptionsoff
  \newpage
\fi



 \bibliographystyle{IEEEtran}
 \bibliography{Ref}
%

%








\end{document}